\documentclass{aa}

\usepackage[varg]{txfonts}
\usepackage{graphicx}
\usepackage{url}
\usepackage{natbib}
\usepackage{color}
\usepackage{multirow}

\bibpunct{(}{)}{;}{a}{}{,} 
\usepackage{amstext}

\begin{document}


\title{Stellar populations of galaxies in the ALHAMBRA survey up to $z \sim 1$\thanks{Based on observations collected at the Centro Astron\'omico Hispano Alem\'an (CAHA) at Calar Alto, operated jointly by the Max-Planck Institut f\"ur Astronomie and the Instituto de Astrof\'isica de Andaluc\'ia (CSIC)}}
\subtitle{IV. Properties of quiescent galaxies on the stellar mass--size plane}


\author{L.~A.~D\'iaz-Garc\'ia\inst{\ref{a1},\ref{a0}}
\and A.~J.~Cenarro\inst{\ref{a1a}}
\and C.~L\'opez-Sanjuan\inst{\ref{a1a}}
\and L.~Peralta de Arriba\inst{\ref{a01}}
\and I.~Ferreras\inst{\ref{a2}}
\and M.~Cervi\~no\inst{\ref{a3},\ref{a4},\ref{a5}}
\and I.~M\'arquez\inst{\ref{a3}}
\and J.~Masegosa\inst{\ref{a3}}
\and A.~del Olmo\inst{\ref{a3}}
\and J.~Perea\inst{\ref{a3}}
}

%
\institute{Centro de Estudios de F\'isica del Cosmos de Arag\'on (CEFCA), Plaza San Juan 1, Floor 2, E--44001 Teruel, Spain\label{a1}\\  \email{ladiaz@asiaa.sinica.edu.tw}
\and Academia Sinica Institute of Astronomy \& Astrophysics (ASIAA), 11F of Astronomy-Mathematics Building, AS/NTU, No.~1, Section 4, Roosevelt Road, Taipei 10617, Taiwan\label{a0}
\and Centro de Estudios de F\'isica del Cosmos de Arag\'on (CEFCA) - Unidad Asociada al CSIC, Plaza San Juan 1, Floor 2, E--44001 Teruel, Spain\label{a1a}
\and Institute of Astronomy, University of Cambridge, Madingley Road, Cambridge CB3 0HA, UK\label{a01}
\and Mullard Space Science Laboratory, University College London, Holmbury St Mary, Dorking, Surrey RH5 6NT, United Kingdom\label{a2}
\and IAA-CSIC, Glorieta de la Astronom\'ia s/n, 18008 Granada, Spain\label{a3}
\and Instituto de Astrof\'isica de Canarias, V\'ia L\'actea s/n, 38200 La Laguna, Tenerife, Spain\label{a4}
\and Depto.~Astrof\'isica, Centro de Astrobiolog\'ia (INTA-CSIC), ESAC campus, Camino Bajo del Castillo s/n, E-28692 Villanueva de la Ca\~{n}ada, Spain\label{a5}
}

\date{Received ? / Accepted ?}

\abstract{}  
         {We perform a comprehensive study of the stellar population properties (formation epoch, age, metallicity, and extinction) of quiescent galaxies as a function of size and stellar mass to constrain the physical mechanism governing the stellar mass assembly and the likely evolutive scenarios that explain their growth in size.}            
         {After selecting all the quiescent galaxies from the ALHAMBRA survey by the dust-corrected stellar mass-colour diagram, we built a shared sample of $\sim850$ quiescent galaxies with reliable measurements of sizes from the HST. This sample is complete in stellar mass and luminosity, $I\le23$. The stellar population properties were retrieved using the fitting code for spectral energy distributions called MUlti-Filter FITting for stellar population diagnostics (MUFFIT) with various sets of composite stellar population models. Age, formation epoch, metallicity, and extinction were studied on the stellar mass--size plane as function of size through a Monte Carlo approach. This accounted for uncertainties and degeneracy effects amongst stellar population properties.}     
         {The stellar population properties of quiescent galaxies and their stellar mass and size since $z\sim1$ are correlated. At fixed stellar mass, the more compact the quiescent galaxy, the older and richer in metals it is ($1$~Gyr and $0.1$~dex, respectively). In addition, more compact galaxies may present slight lower extinctions than their more extended counterparts at the same stellar mass ($<0.1$~mag). By means of studying constant regions of stellar population properties across the stellar mass-size plane, we obtained empirical relations to constrain the physical mechanism that governs the stellar mass assembly of the form $M_\star \propto r_\mathrm{c}^\alpha$, where $\alpha$ amounts to $0.50$--$0.55 \pm 0.09$. There are indications that support the idea that the velocity dispersion is tightly correlated with the stellar content of galaxies. The mechanisms driving the evolution of stellar populations can therefore be partly linked to the dynamical properties of galaxies, along with their gravitational potential.}     
         {}  
 

\titlerunning{Stellar populations of quiescent galaxies on the stellar mass--size plane}

\authorrunning{L.~A.~D\'iaz-Garc\'ia et al.}

\maketitle



\section{Introduction}\label{sec:introduction}

During the past decades, tight correlations between the stellar mass of a galaxy and its stellar population properties since moderate redshifts have frequently been found \citep[e.g.][]{Ferreras2000,Kauffmann2003, Gallazzi2005, Thomas2005, SanchezBlazquez2006, Jimenez2007, Kaviraj2007, Panter2008, Vergani2008, Ferreras2009, delaRosa2011, Jorgensen2013,Jorgensen2014,Jorgensen2017,Peng2015,DiazGarcia2017a,DiazGarcia2017b}, where the environment can also be involved in its assembly  \citep[e.g.][]{Thomas2005,Ferreras2006,SanchezBlazquez2006,Rogers2010,LaBarbera2014,Mcdermid2015}. However, some authors concluded that other galaxy properties, such as the stellar surface density \citep[][]{Kauffmann2003,Franx2008} or the velocity dispersion \citep{Trager2000,Gallazzi2006,Graves2010,Cappellari2013}, strongly correlate with the formation and evolution of the stellar content in galaxies, even more than the proper stellar mass.

On the other hand, the sizes of local galaxies clearly correlate with their stellar masses and luminosities \citep[see e.g.][]{Kormendy1977,Blanton2003,Shen2003}, with a tight dependence on morphological aspects and colours (early-type and red galaxies are typically more compact or denser than the late-type and blue galaxies at fixed stellar mass or luminosity). More recently, many studies unveiled striking evidence in favour of a continuous and generalised increase in size of both spheroid-like and quiescent and late-type and star-forming galaxies across cosmic time \citep[e.g.][]{Trujillo2004, Daddi2005,Mcintosh2005,Trujillo2006,Toft2007,Trujillo2007,Zirm2007,Buitrago2008,vanDokkum2008,Damjanov2011,Newman2012,vanderWel2014}. In particular, since $z\sim2$ ($z\sim1$) massive spheroid-like and quiescent galaxies have rapidly increased in size by a factor of $\sim4$ ($1.5$--$2$) up to the present. However, the main mechanism that causes this fast growth in size is not yet completely clear. 

First attempts to distinguish why galaxies currently exhibit larger sizes have proposed that the influence of active galactic nuceli (AGNs) might play a role. This scenario, commonly referred to as the `puffing-up' scenario \citep{Fan2008, Fan2010, Damjanov2009}, suggested that feedback from AGNs or quasars would remove cold gas from the inner regions of the galaxy and in this way redistribute the stellar content of the inner regions on a timescale of $\sim2$~Gyr. As a result of this mechanism, more extended galaxies would present older ages in the local Universe and the dispersion of the stellar mass--size relation would increase across cosmic time \citep{Fan2010}. Nevertheless, many observational studies have described the opposite case because they found no significant increase in the dispersion of this relation, and more compact galaxies show features that are expected for older stellar populations \citep[e.g.][]{Cenarro2009, Shankar2009,Trujillo2009,Trujillo2011,Mcdermid2015,PeraltaArriba2015,Gargiulo2017}. For these reasons, this mechanism has mostly been discarded as an explanation of the extended growth in size of galaxies. 

Alternatively, mergers were proposed as an efficient mechanism to produce a generalised growth in size \citep{Naab2009}. Under this scenario, galaxies firstly formed their compact and dense cores, which could be the result of mergers between gas-rich discs \citep[which yields compact starbursts of small radii,][]{Hopkins2008} or of the accretion of cold streams \citep[which forms compact massive bulges and suppresses star formation,][]{Keres2005,Dekel2009}. These cores would be the so-called red nuggets observed at $z > 2$ \citep[][]{Damjanov2009,delaRosa2016}. When the core is assembled, galaxies would continue their assembly by a continuous fall of pieces at lower redshifts through mergers (with ex situ stellar populations). In this way, the surroundings of dense cores would be populated in an `inside-out' formation scenario \citep[e.g.][]{delaRosa2016}. This scenario would be mainly driven by minor mergers on parabolic orbits \citep{Khochfar2006a,Khochfar2006b,Bezanson2009,Hopkins2009,Naab2009,Trujillo2011} because there are not enough major mergers to reproduce the evolution in size observed since $z\sim1$ \citep{Bundy2009,deRavel2009,LopezSanjuan2010,LopezSanjuan2012}. Because merger histories do not preferentially involve compact galaxies \citep[][]{DiazGarcia2013}, the growth in size through mergers would be generalised for all the galaxies in the stellar mass--size plane. Under these assumptions, the number of compact galaxies would be reduced at longer cosmic times \citep[][]{Cassata2013,Quilis2013,Trujillo2014,vanderWel2014}. However, the variation with redshift of the number of massive galaxies, with a prominent growth in size, is still a matter of debate. Contradictory studies point out that the number of compact galaxies has remained almost constant since intermediate redshifts \citep{Saracco2010,Damjanov2014,Damjanov2015,Gargiulo2017} or experienced only a slight decrease in number \citep{Valentinuzzi2010,Poggianti2013}. A reliable estimation of the evolution in number of compact and massive galaxies across cosmic time may help to discern whether mergers should be considered as an important mechanism for the growth in size of quiescent or spheroid-like galaxies.

In the past years, the progenitor bias \citep[][]{vanDokkum2001} has received more attention. According to this, samples of galaxies at high redshift are biased sub-sets of the nearby counterparts because the samples only include the oldest members of the current distributions. Under this assumption, the first galaxies that formed in the earliest epochs of the Universe were also the densest because they resided in denser halos. At the same time, they would evolve and quench their star-formation processes earlier. Any galaxy formed at later epochs will therefore be larger and evolve later until it reached enough stellar mass for no more star-forming processes to be supported. This would imply that less dense quiescent galaxies (less compact or extended) are also younger because they reach a state of quenching in later epochs \citep{Valentinuzzi2010,Carollo2013,Belli2015}. For the quiescent case, after these galaxies quenched their main star formation processes, they would lie on the upper parts of the stellar mass--size plane, yielding a growth of the median size according to the observations of the mass--size relation of star-forming and quiescent galaxies \citep[e.g.][]{Trujillo2007,Zirm2007,Buitrago2008,Newman2012,vanderWel2014}. At the same time, this would imply that the number density of compact galaxies would be constant, or at least this would suffer mild and increasing modifications across cosmic time. Owing to the progenitor bias, we would expect a correlation between the size of a galaxy and their stellar content age, where denser galaxies exhibit older ages \citep{Shankar2009,Poggianti2013,Fagioli2016,Gargiulo2017,Williams2017,Wu2018,Damjanov2019} or large quiescent galaxies will reach the red sequence later than their compact counterparts \citep{Belli2015}. 

A detailed analysis of stellar populations of compact and extended quiescent galaxies will allow us to shed light on the scenarios or mechanisms related to the prominent growth in size that these galaxies have undergone since high redshift. If extended galaxies would be found to be older than their compact counterparts, it would favour an inside-out formation, whereas the opposite case could be consistent with mergers and progenitor bias. Probing the stellar content of these galaxies in a long period of cosmic time will also provide valuable information with which more complex scenarios could be tested that include not one unique mechanism acting in favour of a growth in size. For instance, \citet{Belli2015} showed that the progenitor bias contribution at $1 < z < 1.6$ can only explain half of the average size evolution that quiescent galaxies exhibit during $1.25 < z < 2$. In addition, for the generalised evolution of the median metallicity of massive quiescent galaxies reported by \citet[][]{DiazGarcia2017b}, mergers would be necessary, as would the progenitor bias to interpret the results properly.

This work is part of a series of papers in which our goal is to improve our understanding of quiescent galaxies since $z\sim1$, with the ultimate goal of providing a general picture of the formation and evolution of quiescent galaxies, for which we make use of multiple observables (e.g.~number densities, stellar population properties, and sizes). In this work, we present a comprehensive study of the stellar populations of a set of quiescent galaxies as a function of their size, mass, and redshift, in pursuit of finding the most plausible evolution scenarios that can explain their growth in size.

Overall, the structure of this work is as follows. We define the quiescent sample with reliable sizes and stellar population properties in Sect.~\ref{sec:sizes_sample}. In Sect.~\ref{sec:sp_size} we explore the correlations between the sizes and age, formation epoch, metallicity, and extinction of quiescent galaxies. An empirical expression for the driver of the stellar content of galaxies is obtained in Sect.~\ref{sec:driver}. Our results are discussed and compared with previous work in Sects.~\ref{sec:discussion_grow} and \ref{sec:sizes_previous}, respectively. A brief summary of this research is presented in Sect.~\ref{sec:sizes_conclusions}.

We adopt a lambda cold dark matter ($\Lambda$CDM) cosmology with $H_0 = 71$~km~s$^{-1}$, $\Omega_\mathrm{M}=0.27$, and $\Omega_\mathrm{\Lambda}=0.73$ throughout. Stellar masses are quoted in solar mass units $\mathrm{[M_\sun]}$ and magnitudes in the AB-system \citep{Oke1983}. We assume \citet{Chabrier2003} and Kroupa universal \citep[][]{Kroupa2001} initial stellar mass functions (IMF, more details in Sect.~\ref{sec:data_sp}).


\section{Sample of quiescent galaxies with reliable sizes}\label{sec:sizes_sample}

The reference catalogue for this work is the catalogue of quiescent galaxies published by \citet{DiazGarcia2017a}. This catalogue includes $\sim8\,500$ quiescent galaxies from the Advanced Large Homogeneous Area Medium Band Redshift Astronomical survey \footnote{\url{http://www.alhambrasurvey.com}} \citep[ALHAMBRA, ][]{Moles2008} at $0.1 \le z \le 1.1$, whose stellar population properties were constrained through spectral energy distribution (SED) fitting techniques. These stellar population properties were studied in detail by \citet{DiazGarcia2017b}. This sample of galaxies was defined to be complete in stellar mass ($95$~\% complete at any redshift bin) and luminosity ($I\le 23$), where a dust-corrected stellar mass-colour diagram (or MCDE) was used to minimise the contamination of dusty star-forming galaxies. In order to assess likely systematics on the stellar population properties owing to the use of population synthesis models, two sets of simple stellar populations (SSP) models were used: the set described by \citet[][hereafter BC03]{Bruzual2003} and that of \citet[][EMILES]{Vazdekis2016}. The method for determining stellar population properties is briefly presented in Sect.~\ref{sec:data_sp}. As our goal is to shed light on potential correlations between the stellar populations of galaxies and their sizes, we complement this catalogue with reliable size measurements of galaxies. Further details are provided in Sect.~\ref{sec:hst_sizes}.

\begin{table*}
\caption{Number of quiescent galaxies with reliable sizes obtained from ACS/HST fields in common with ALHAMBRA for a stellar mass completeness level of $95$~\%.}
\label{tab:acs}
\centering
\begin{tabular}{ccccccc}
\hline\hline
&&&&&& \\
ACS field & ALHAMBRA field &  & Number &  & ACS Filter & ACS pixel scale\\
\cline{3-5} \\
 & & EMILES & EMILES & BC03 & & [pixel$^{-1}$]\\
 & & (Padova00) & (BaSTI) &  & & \\
&&&&&& \\
\hline
&&&&&& \\
COSMOS$^1$ & ALH-4 & $683$ & $677$ & $715$ & $F814W$ & $0 \farcs 05$ \\
HDF-N$^{2,3}$ & ALH-5 & $44$ & $42$ & $41$ & $F775W$ & $0 \farcs 03$ \\
AEGIS$^4$ & ALH-6 & $101$ & $111$ & $114$ & $F814W$ & $0 \farcs 03$ \\

\multirow{2}{*}{Total} & & \multirow{2}{*}{$828$} & \multirow{2}{*}{$830$} & \multirow{2}{*}{$870$} & & \\
&&&&&& \\
\hline
\end{tabular}
\tablefoot{(1) \citet{Scoville2007}; (2) \citet{Dickinson2003}; (3) \citet{Giavalisco2004}; (4) \citet{Davis2007}.}
\end{table*}


\subsection{Stellar population properties of quiescent galaxies from the ALHAMBRA survey}\label{sec:data_sp}

The ALHAMBRA survey provides flux in $23$ photometric bands\footnote{\url{http://svo2.cab.inta-csic.es/theory/fps3/}}, $20$ in the optical range \citep[$\lambda\lambda\ 3500$--$9700$~\AA, with a full width at half-maximum of $\sim300$~\AA, see details in][]{Aparicio2010} and the near-infrared bands $J$, $H$, and $K_\mathrm{s}$ \citep[][]{Cristobal2009}. This survey was acquired with the $3.5$ m telescope of the Calar Alto Observatory\footnote{\url{http://www.caha.es}} (CAHA), covering a current effective area of $\sim2.8$~deg$^2$ in $7$ non-contiguous fields in the northern hemisphere. The reference catalogue of photometry for this work was the ALHAMBRA Gold catalogue\footnote{\url{http://cosmo.iaa.es/content/alhambra-gold-catalog}} \citep{Molino2014}, which contains $\sim95\,000$ galaxies down to $I=23$ with accurate photo-redshifts (hereafter photo-$z$) constraints \citep[$\sigma_z \sim 0.012$, further details in][]{Molino2014}. This catalogue provides point spread function (PSF) corrected \citep{Coe2006} and non-fixed aperture photometry, which is ideally suited for stellar population studies, including the total flux emitted by galaxies without biasing their outer parts.

The stellar population properties of quiescent galaxies \citep[age, metallicity, extinction, stellar mass, photo~$z$, and rest-frame luminosities; see][]{DiazGarcia2017a} were determined by the use of the SED-fitting code called MUlti-Filter FITting for stellar population diagnostics \citep[MUFFIT, ][]{DiazGarcia2015}. To assess potential systematics caused by the varying model prescriptions in these sets, MUFFIT was fed with two independent sets of SSP models to construct a set of composite models of stellar populations for each of them. These composite models consist of a mixture of two individual SSP models, also referred to as two burst formation models. Consequently, the star formation history of quiescent galaxies is approached by one old and one young component, which has proven to reproduce the colour of an underlying red population plus less massive and later events of star formation \citep[e.g.][]{Ferreras2000,Kaviraj2007,Rogers2010}. For this work, we took the SSP models of BC03 and EMILES to build our composite models of stellar population and to subsequently constrain the stellar population properties of quiescent galaxies thorugh SED-fitting. For BC03, 19 ages from $0.06$ to $14$~Gyr, metallicity values of $[\mathrm{M/H}]=-1.65$, $-0.64$, $-0.33$, $0.09$, $0.55$~dex (Padova1994 stellar tracks), and the \citet{Chabrier2003} IMF were used to constrain stellar population properties. Regarding EMILES, the two sets of theoretical isochrones were taken into account: the scaled-solar isochrones of \citet[][hereafter Padova00]{Girardi2000}, and those of \citet[][BaSTI in the following]{Pietrinferni2004}. Twenty-two ages in the range $0.05$--$14$~Gyr and metallicities of $[\mathrm{M/H}]=-1.31$, $-0.71$, $-0.40$, $0.00$, $0.22$ for Padova00 and $[\mathrm{M/H}]=-1.26$, $-0.96$, $-0.66$, $-0.35$, $0.06$, $0.26$, $0.40$ for BaSTI were selected to perform the SED-fitting analysis, both with a Kroupa Universal IMF \citep[][]{Kroupa2001}. For both BC03 and EMILES, the extinction law of \citet{Fitzpatrick1999} was assumed for dust reddening with values in the range $A_V=0.0$--$3.1$ and taking a fixed ratio of $R_V=3.1$. We note that \citet{DiazGarcia2017a} stated that stellar masses computed using EMILES SSP models are $\sim0.10$~dex more massive than those obtained with the BC03 models. For this reason, stellar mass completeness limits of EMILES and BC03 differ at any redshift in all the sections of this work.

Throughout this work, the mass-weighted age and metallicity ($\mathrm{Age_M}$ and $\mathrm{[M/H]_M}$, respectively) are preferred to the luminosity age and metallicity. The mass-weighted parameters are more representative of the total stellar content of the galaxy, and they are not linked to a defined luminosity weight, which may differ in different works. Look-back times, $t_\mathrm{LB}$, were established following the recipes of \citet[][]{Hogg1999}. Hereafter, we define the formation epoch as the addition of mass-weighted ages and look-back times, meaning $\mathrm{Age_M}+t_\mathrm{LB}$. 


\subsection{Retrieval of sizes of quiescent galaxies}\label{sec:hst_sizes}

The measurement of galaxy sizes from ground-based surveys, for instance ALHAMBRA, is biased to galaxies with sizes larger than the spatial resolution of the survey. In addition, the evolution of the stellar mass--size relation with redshift complicates a reliable and non-biased measurement of galaxy sizes. To overcome this drawback, we took advantage of the ALHAMBRA fields that overlap with fields of the \textit{Hubble} Space Telescope (HST), where these limitations do not exist at the ALHAMBRA redshifts.

From the ALHAMBRA galaxy sample presented in \citet{DiazGarcia2017a}, we constructed a sub-set of shared quiescent galaxies with accurate effective radius measurements derived from HST data. Size measurements were retrieved from the Advanced Camera for Surveys (ACS) general catalogue of structural parameters \citep[][hereafter ACS catalogue]{Griffith2012}. The ACS catalogue was constructed using the GALAPAGOS\footnote{Galaxy Analysis over Large Areas: Parameter Assessment by GALFITting Objects from SExtractor} code \citep[][]{Haussler2011}, which includes both SExtractor \citep[][]{Bertin1996} and GALFIT\footnote{A two-dimensional (2D) fitting algorithm} \citep[][]{Peng2002}, to model and measure structural parameters of each source in the ACS catalogue assuming a \citet[][]{Sersic1968} profile. In particular, sizes were computed using as initial values to iterate the FLUX\_RADIUS, $R_\mathrm{f}$, and MAG\_BEST provided by SExtractor and the formula $r_\mathrm{e} = 0.162\ R_\mathrm{f}^{1.87}$, which was derived through simulations \citep[further details in][]{Griffith2012}. In the following, our reference measurement for size is the circularised effective radius, $r_\mathrm{c}$, instead of the provided effective radius, $r_\mathrm{e}$, which encloses half of the total flux. It is formally expressed as $r_\mathrm{c} = r_\mathrm{e} \sqrt{a_1/a_0}$, where $a_1$ and $a_0$ are the semi-major and semi-minor axes, respectively. The projected sizes from \citet{Griffith2012} were converted into physical units and circularised using the axis ratio provided by the ACS catalogue ($a_1/a_0$, column \texttt{BA\_GALFIT\_HI}). 

After cross-correlating all the quiescent galaxies from the ALHAMBRA survey (fields 4, 5, and 6) at a stellar mass completeness level of\ $95~\%$ with the ACS catalogue, we finally obtained a common sample of $870$, $830$, and $828$ quiescent galaxies for BC03, EMILES+BaSTI, and EMILES+Padova00, respectively. There is a $90$~\% fraction of galaxies in common in the two EMILES and BC03 samples. Owing to a larger overlapping area with ALHAMBRA, most of structural parameters of quiescent galaxies in this study come from the Cosmological Evolution Survey field \citep[COSMOS, ][]{Scoville2007}. In the following, our reference bands for morphological parameters (including apparent size) in the ACS catalogue are $F814W$, for COSMOS and the All-wavelength Extended Groth Strip International Survey \citep[AEGIS][]{Davis2007}, and $F775W$ for the Hubble Deep Field North \citep[HDF-N][]{Dickinson2003,Giavalisco2004}. In Table~\ref{tab:acs} we show a brief summary of the characteristics of this sub-sample. We stated the redshift upper limit of this work at $z=0.9$ because the sample at $z>0.9$  is constrained to quiescent galaxies with $\log_{10}M_\star \gtrsim 11.2$ and only a few sources have ACS size measurements.


\subsection{Correction for colour gradients}\label{sec:hst_corr}

The presence and evolution of colour gradients in galaxies has been extensively studied since moderate redshifts \citep[e.g.][]{Ferreras2005,Szomoru2011,vanderWel2014}. Colour gradients affect the measurement of structural parameters such as half-light radius, which decreases at increasing wavelength \citep[see e.g.][]{Kelvin2012,Wuyts2012}. This is especially relevant in studies involving galaxies at different redshift with a few bands to determine structural parameters. According to this, the effective radius obtained by the $F814W$ band at $z=0.63$ and by $F775W$ at $z=0.55$ corresponds to the effective radii at a rest-frame wavelength of $5\,000$~\AA. 

As our sub-sample extends from $z=0.1$ to $z=0.9$, we mitigated this passband-shifting effect by converting the size measurements obtained from the ACS $F814W$ and $F775W$ images into a common rest-frame wavelength of $5\,000$~\AA. Following \citet{vanderWel2014}, $r_\mathrm{c}$ at a rest-frame wavelength of $5\,000$~\AA\ for a galaxy at redshift $z$ is obtained through the expression
\begin{equation}\label{eq:col_grad}
r_\mathrm{c} = r_\mathrm{c,F}\left(\frac{1+z}{1+z_\mathrm{p}}\right) ^{\frac{\Delta \log_{10}r_\mathrm{c}}{\Delta \log_{10}\lambda}}\ ,
\end{equation}
where F denotes the bands $F814W$ (COSMOS and AEGIS) and $F775W$ (HDF-N), $z_\mathrm{p}$ is the pivot wavelength ($0.63$ and $0.55$ for $F814W$ and $F775W$ bands, respectively), and $\frac{\Delta \log_{10}r_\mathrm{c}}{\Delta \log_{10}\lambda}$ is the size gradient as a function of wavelength. For the size gradient, we assumed the value reported by \citet{vanderWel2014} for quiescent galaxies, $\Delta \log_{10}r_\mathrm{c} / \Delta \log_{10}\lambda=-0.25$, which is also in agreement with \citet[][]{Guo2011} and \citet[][]{Kelvin2012}. Negative values for $\frac{\Delta \log_{10}r_\mathrm{c}}{\Delta \log_{10}\lambda}$ imply that quiescent galaxies are more compact at longer wavelengths. The size corrections introduced by Eq.~\ref{eq:col_grad} are mild and do not exceed a $10$~\% fraction with respect the observed values, meaning $0.95 < r_\mathrm{c}/r_\mathrm{c,F} < 1.1$.

\begin{figure*}
\centering
\includegraphics[width=17cm,clip=True]{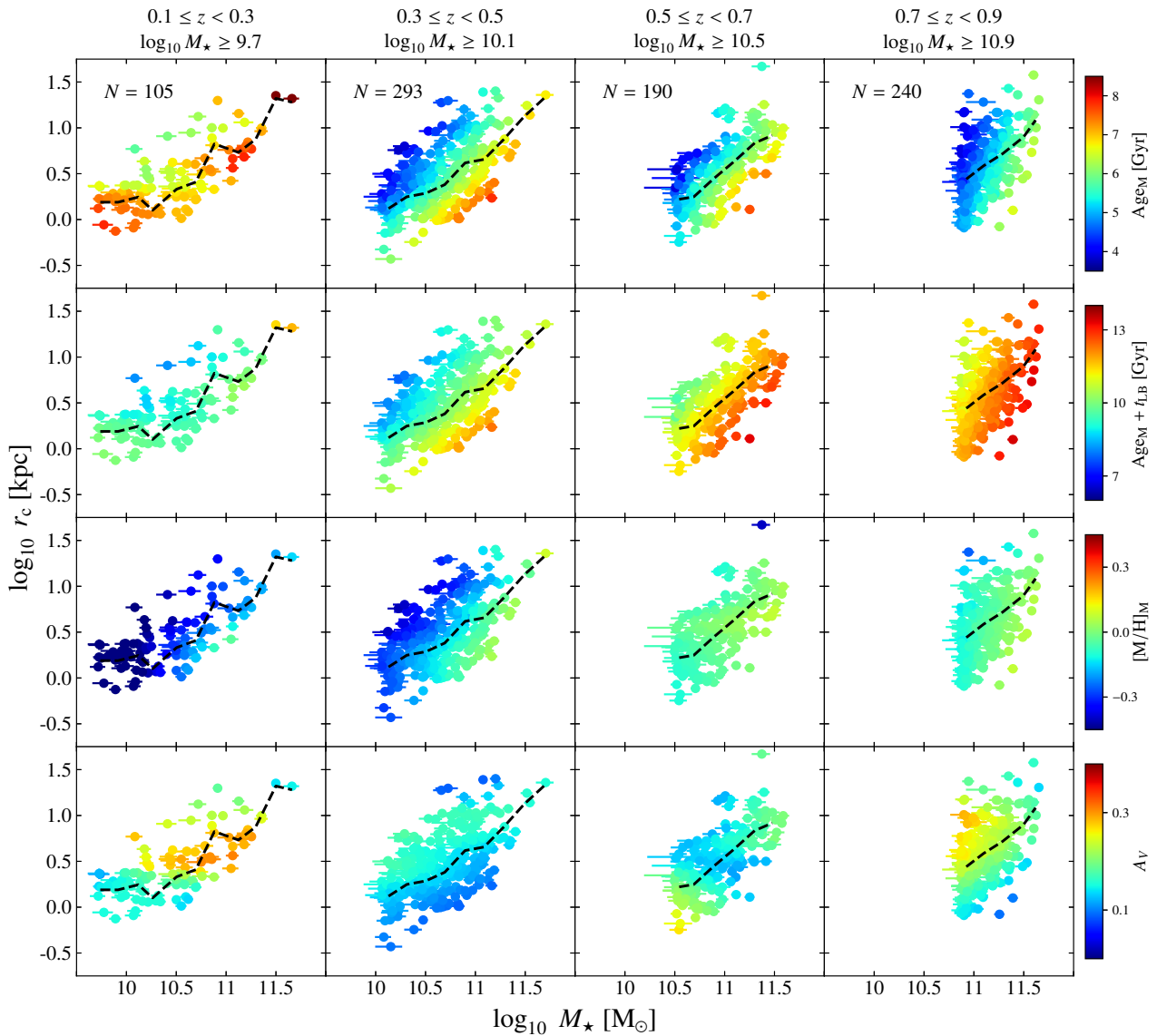}
\caption{From top to bottom, mass-weighted age, formation epoch, metallicity, and extinction of quiescent galaxies using EMILES+Padova00 models plotted on the circularised radius vs.~stellar mass plane down to $z=0.9$. All values are colour-coded as indicated by each colour bar and averaged with the LOESS method (for more details, see Sect.~\ref{sec:sp_size}). The dashed line illustrates the median circularised radius. The number of galaxies in each panel is pointed out in the first row.}
\label{fig:sizes_padova}
\end{figure*}

\begin{figure*}
\centering
\includegraphics[width=17cm,clip=True]{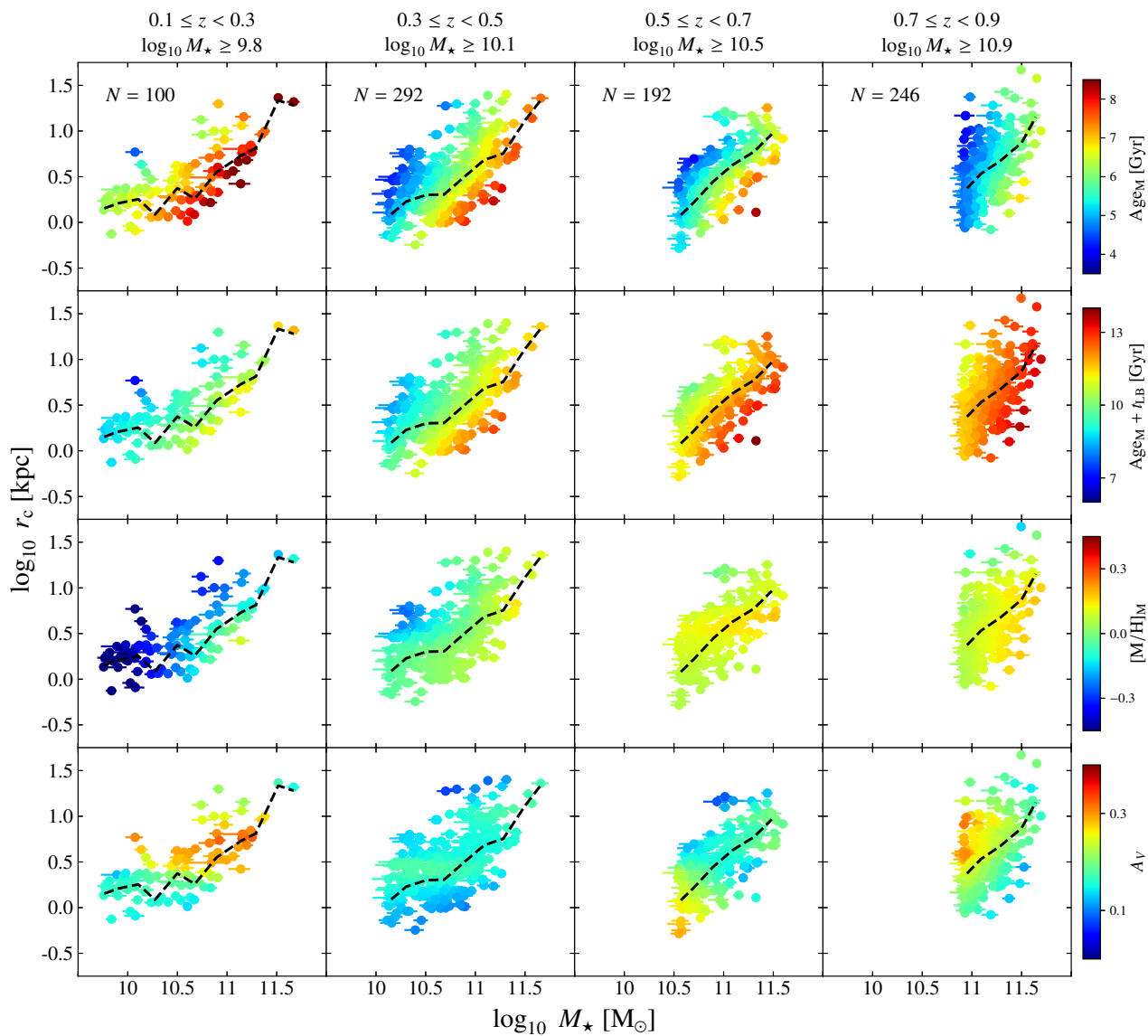}
\caption{Same as Fig.~\ref{fig:sizes_padova}, but for EMILES+BaSTI SSP models.}
\label{fig:sizes_basti}
\end{figure*}

\begin{figure*}
\centering
\includegraphics[width=17cm,clip=True]{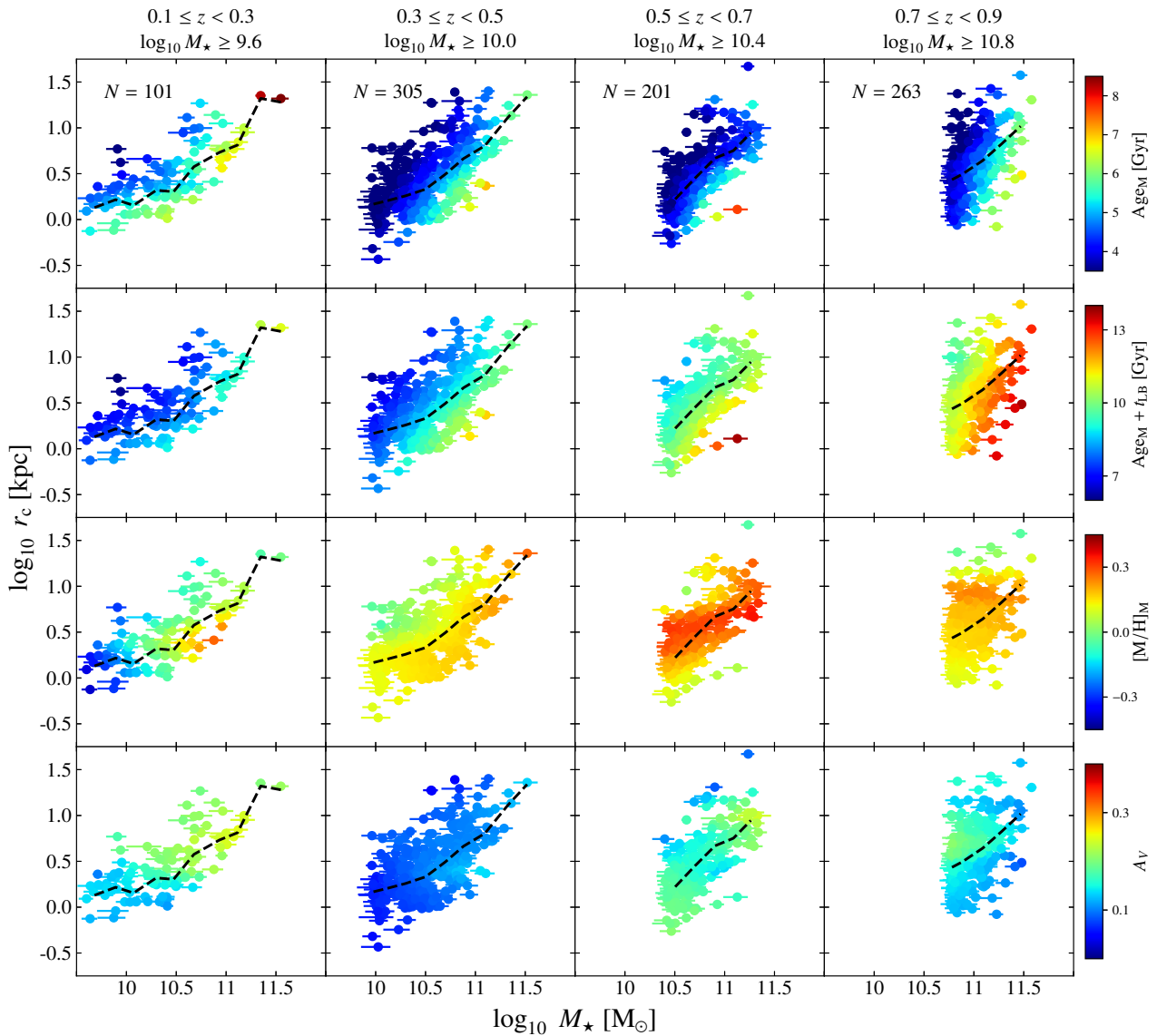}
\caption{Same as Fig.~\ref{fig:sizes_padova}, but for BC03 SSP models.}
\label{fig:sizes_bc03}
\end{figure*}

\begin{figure*}
\centering
\includegraphics[width=17cm,trim= 0 1.86cm 0 0,clip=True]{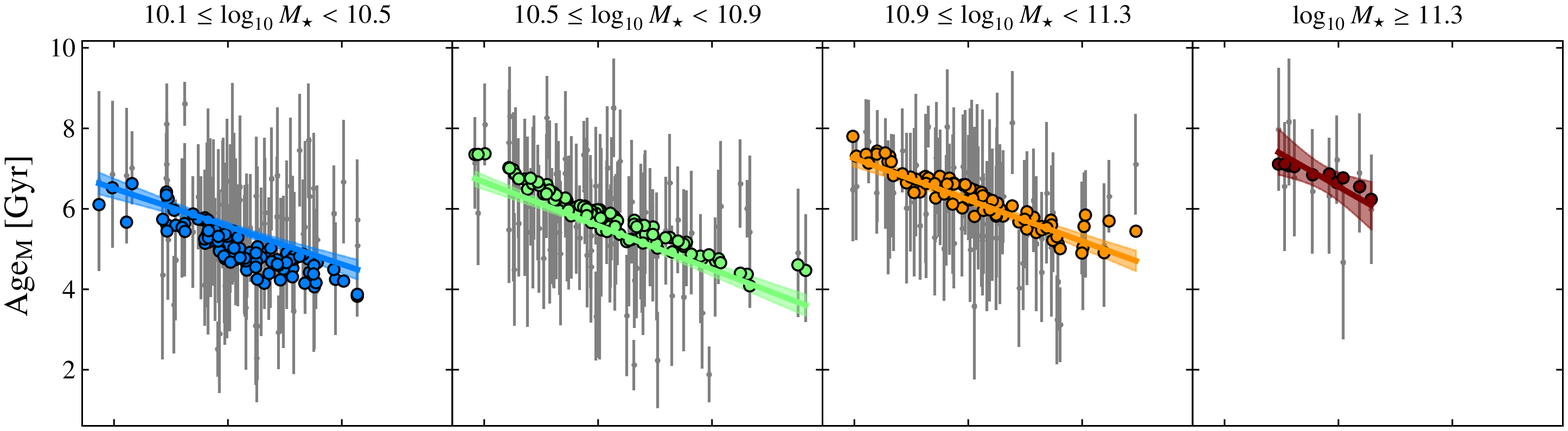}
\includegraphics[width=17cm,trim= 0 1.86cm 0 1.007cm,clip=True]{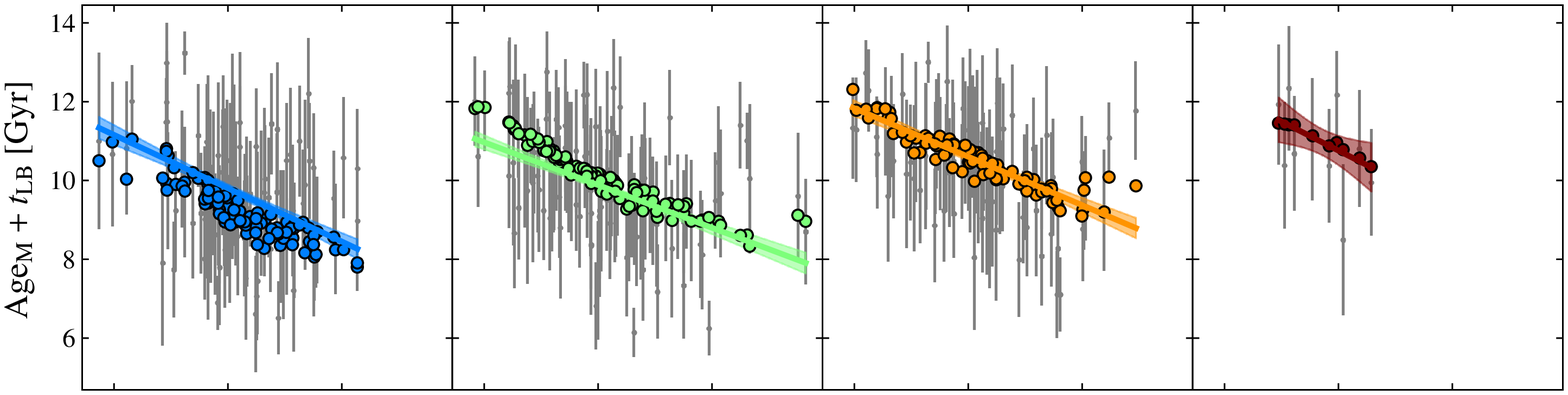}
\includegraphics[width=17cm,trim= 0 1.86cm 0 1.007cm,clip=True]{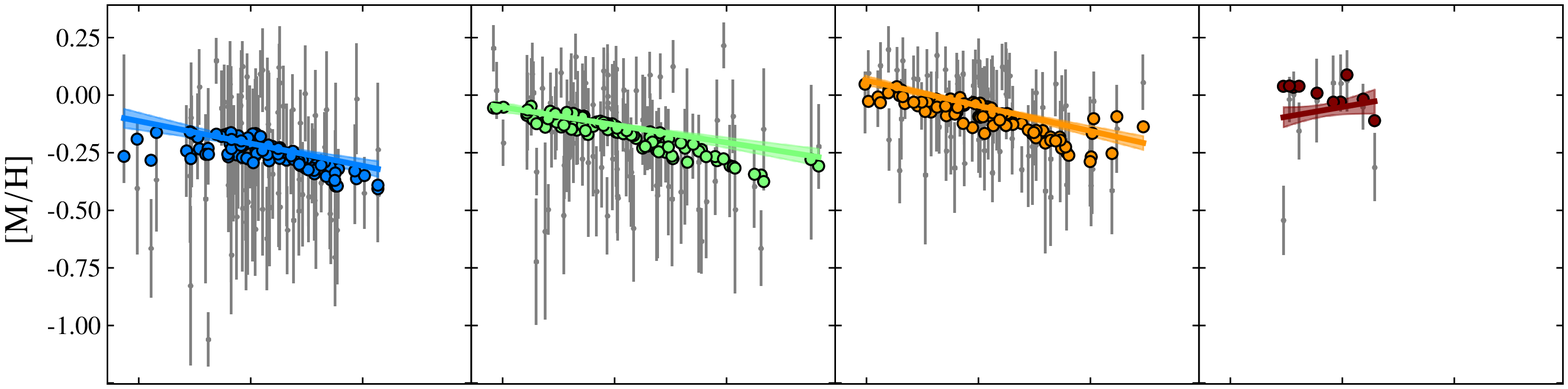}
\includegraphics[width=17cm,trim= 0 0 0 1.007cm,clip=True]{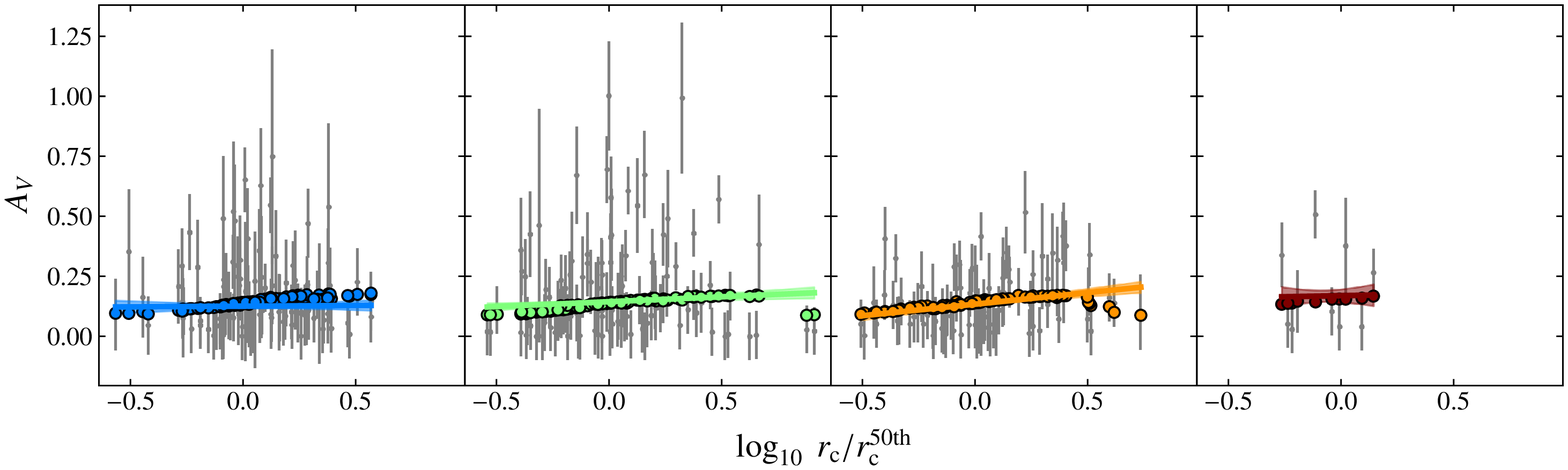}
\caption{Individual values of stellar population parameters of quiescent galaxies as a function of their size and median circularised radius ($r_\mathrm{c}$ and $r_\mathrm{c}^\mathrm{50th}$, respectively) at $0.3\le z < 0.5$. From top to bottom, we show the mass-weighted age, formation epoch, metallicity, and extinction for EMILES+Padova00 models. Each panel comprises a different stellar mass range. Vertical bars enclose the 1~$\sigma$ uncertainty level of individual values. Solid lines show the linear least-squares fitting to the distribution of values in each panel (see Eq.~\ref{eq:size_sp_par}), whereas the shaded region is its 1~$\sigma$ level uncertainty. Coloured markers illustrate the smoothed values obtained with the LOESS method (see Fig.~\ref{fig:sizes_padova}).}
\label{fig:noloess}
\end{figure*}


\section{Correlations between size and stellar population parameters of quiescent galaxies}\label{sec:sp_size}

In this section, we unveil the correlations that exist between sizes of galaxies and their stellar contents. These results might be explored to shed light on the growth in size mechanisms of quiescent galaxies since $z=1$. 

To study the distribution of the stellar population parameters on the stellar mass--size plane, we took advantage of the bidimensional and locally weighted regression method, or LOESS \citep[][]{Cleveland1979,Cleveland1988}. This method finds a non-parametric plane to reveal trends on the stellar mass--size plane through the distribution of points (stellar population parameters for this study), which minimises the uncertainty effects in the diagrams by a regression technique. In particular, we used the \texttt{Python} implementation of the LOESS method published by \citet[][]{Cappellari2013b}. During the LOESS process, we set a regularisation factor or smoothness of $f=0.6$ for all the solutions retrieved during the Monte Carlo approach. In this way, we included correlations and degeneracies between parameters. This is necessary because both sizes and stellar masses also depend on the photo-~\textit{z} determination, as do the remaining stellar population parameters explored here. We report mass-weighted age and formation epoch (Sect.~\ref{sec:size_age}), metallicity (Sect.~\ref{sec:size_feh}), and extinction (Sect.~\ref{sec:size_av}). The results of the smoothed distributions of the stellar population properties within the stellar mass--size plane obtained with the LOESS method are shown in Figs.~\ref{fig:sizes_padova} to \ref{fig:sizes_bc03}. Moreover, in Fig.~\ref{fig:noloess} we illustrate the variation in individual stellar population properties and uncertainties provided by MUFFIT as a function of the size, as well as the values retrieved for the same quiescent galaxies after the LOESS smoothing.

In addition, we are able to quantify the variation in different stellar population parameters with size. This relation is expressed as a function of the median circularised radius, $r_\mathrm{c}^\mathrm{50th}$, observed at each redshift and stellar mass. We assumed a linear dependence on size of the form
\begin{equation}\label{eq:size_sp_par}
p(z,M_\star,r_\mathrm{c}) = \log_{10}\left(r_\mathrm{c}/r_\mathrm{c}^\mathrm{50th}\right) \cdot \bar{m}_p(z,M_\star) + \bar{n}_p(z,M_\star)\ ,
\end{equation}
where $p$ represents the stellar population parameter (mass-weighted age, formation epoch, metallicity, and extinction), $\bar{m}_p$ the slope or the correlation with size, and $\bar{n}_p$ the intercept. The linear fitting was performed with the original values provided by MUFFIT (see Fig.~\ref{fig:noloess}) at different redshift and stellar mass bins. The latter to diminish mass dependence effects and differences in median sizes amongst stellar mass ranges.


\subsection{Age and formation epoch in the stellar mass--size plane}\label{sec:size_age}

The results in Figs.~\ref{fig:sizes_padova} to \ref{fig:sizes_bc03} (see first row panels) show a tight correlation between the size of a quiescent galaxy and its mass-weighted age. This result is confirmed independently of the three sets of SSP models (BC03 and EMILES with isochrones BaSTI and Padova00) we used for the stellar population predictions. We find out a clear trend in which more compact galaxies are systematically older at $\log_{10} M_\star \sim 9.6$, whereas younger quiescent galaxies lie on the upper parts of the stellar mass--size relation. This correlation is extended beyond the nearby Universe up to $z=0.9$. Furthermore, there is no strong evidence for a dependence of the size--age correlation on stellar mass, $\langle \bar{m}_\mathrm{Age_M} \rangle \sim -1.4 \pm 0.2$ at $\log_{10} M_\star \gtrsim 9.6$. In addition, there is no evidence for an evolution of the size--age relation through cosmic time, meaning that the redshift dependence in Eq.~(\ref{eq:size_sp_par}) for ages resides in $\bar{n}_\mathrm{Age_M}$. Consequently, the common age difference between the most compact and most extended galaxies\footnote{Quiescent galaxies with a circularised radius $0.3$~dex smaller and larger than the median size at any redshift and stellar mass, respectively. These limits correspond to the $10$th and $90$th percentiles of the size distribution of quiescent galaxies.} is $\sim1$~Gyr. Even though the stellar mass--size relation is less prominent for quiescent galaxies below $\log_{10} M_\star \lesssim 10$--$10.5$ \citep[e.g.][]{Shen2003,Ferrarese2012,vanderWel2014,Guerou2015,Lange2015}, the size--age correlation is also remarkable below this stellar mass limit.

Similar results were obtained for the mass-weighted formation epoch (see second row panels in Figs.~\ref{fig:sizes_padova} to \ref{fig:sizes_bc03}). The more compact the quiescent galaxy, the older its stellar population, that is, more compact quiescent galaxies were formed at earlier epochs than the extended ones at same stellar mass.

Interestingly, quiescent galaxies that populate the upper part of the stellar mass--size plane exhibit higher dust-corrected luminosities at $2\,800$~\AA\ \citep[also provided by][]{DiazGarcia2017a}. This result is supported by the SSP model sets of BC03 and EMILES and for all the stellar mass ranges explored in this work since $z=0.9$. This luminosity reflects a more massive young component or a later event of star formation activity in extended quiescent galaxies \citep[e.g.][]{Ferreras2000,Rogers2010,Vazdekis2016}, which is reflected in the correlation reported above between size and age and formation epoch.


\subsection{Metallicity as a function of size and stellar mass}\label{sec:size_feh}

In the stellar population predictions of BC03 and EMILES, we also found indications for a correlation between the size of a quiescent galaxy and its metal content (see third row panels in Figs.~\ref{fig:sizes_padova} to \ref{fig:sizes_bc03}). The more compact the quiescent galaxy, the higher its metallicity. For EMILES SSP models, the size--metallicity correlation is clearer than for the BC03 models. In addition, this correlation is observed at least since $z=0.9$. Independently of the stellar mass range, more compact quiescent galaxies are also more metal rich than the extended galaxies of the same stellar mass with a correlation of $\langle \bar{m}_\mathrm{[M/H]_M} \rangle \sim -0.11 \pm 0.02$. Because of the uncertainties and differences amongst the three SSP model predictions, and the limitation in the number of quiescent galaxies with reliable size measurements in ALHAMBRA, we cannot confirm that the correlation size--metallicity varies strongly in the redshift range, nor can we report systematic differences amongst different stellar mass ranges at the same redshift. Our results show that on average and using our previous definition, most compact quiescent galaxies at a fixed stellar mass are $\sim0.07$~dex more metal rich than their extended counterparts.

As revealed by \citet[][]{DiazGarcia2017b}, this sample may be affected by cosmic variance at $0.5 \le z < 0.7$. In addition, when the BC03 SSP models are used for the analysis, \citet[][]{DiazGarcia2017b} showed that the lack of quiescent galaxies at this redshift range was more remarkable in ALHAMBRA. The authors also reported a peak or maximum in metallicity that they did not find with EMILES. This means that the metallicity for BC03 SSP models can be affected by this effect, which may mask the real trend of the size--metallicity at $0.5 \le z < 0.7$.


\subsection{Extinction of quiescent galaxies}\label{sec:size_av}

As highlighted by \citet{DiazGarcia2017b}, there is no large discrepancies among the extinctions of quiescent galaxies (fairly constrained $A_V \lesssim 0.6$) without significant variations with redshift. This a priori limits a remarkable correlation between size and extinction. Despite this, there might be a slight indication that more compact galaxies are compatible with lower levels of extinction in their stellar continuum than the extended quiescent galaxies ($\langle \bar{m}_{A_V} \rangle \sim 0.03 \pm 0.01$, see fourth row panels in Figs.~\ref{fig:sizes_padova} to \ref{fig:sizes_bc03}). Owing to the narrow range of extinction values in these galaxies, the extinction differences between compact and extended quiescent galaxies are below $0.1$ at all the  redshift and stellar mass bins we explored. On average, this difference does not exceed a value of $0.05$. There is no evidence for a dependence of the size--extinction on either redshift or stellar mass.


\section{Main driver of the galaxy stellar content}\label{sec:driver}

Many efforts have been made in the past decades to track the evolution of stellar population properties, as well as to determine the main physical mechanism or driver that governs the stellar mass assembly. The tight correlation between the stellar content of a galaxy and its stellar mass has often been reported. This is usually interpreted as evidence that mass drives the stellar population of galaxies. Nevertheless, many authors reported that the stellar content of galaxies may be driven by other physical mechanisms such as dynamical and/or morphological properties. Our results agree with this fact (see Sect.~\ref{sec:sp_size}), pointing out that properties of stellar populations in galaxies are not only related to the mass, but also to size. By means of studying the correlations between the different stellar population properties across the stellar mass--size plane (Sect.~\ref{sec:sp_size}), we can retrieve empirical relations to set constraints on the mechanisms that govern the stellar mass assembly (Sect.~\ref{sec:driver_sp}). In addition, we are able to explore whether other properties of quiescent galaxies such as the stellar mass--size relation (Sect.~\ref{sec:driver_size}), mass, and luminosity surface densities (Sect.~\ref{sec:driver_density}), morphology (Sect.~\ref{sec:driver_sersic}), and velocity dispersions (Sect.~\ref{sec:driver_velocity}) exhibit correlations across the stellar mass--size plane similar to those obtained for the stellar population properties explored in this work.


\subsection{Empirical constraints on the stellar population driver}\label{sec:driver_sp}

For this aim, we empirically distinguished the regions of the stellar mass--size plane that show similar stellar population parameters. A visual inspection of Figs.~\ref{fig:sizes_padova} to \ref{fig:sizes_bc03} shows that constant formation epochs of quiescent galaxies lie at well-defined positions within the stellar mass--size plane. The formation epoch of a galaxy has been proposed in previous studies as one of the stellar population properties that better correlates with stellar mass \citep[e.g.][]{Thomas2005,Ferreras2009}. Nevertheless, we find that at fixed stellar mass the formation epoch also correlates with size. On the other hand, previous studies have shown evidence for a non-passive evolution of the stellar populations of quiescent galaxies \citep[e.g.][]{Schiavon2006,Gallazzi2014,Siudek2017,DiazGarcia2017b}, which modifies their formation epochs at the same time that galaxies grow in size \citep[][]{Shen2003,Trujillo2007,vanDokkum2008,vanderWel2014}. For these reasons and to unveil correlations with this parameter, we fit the original formation epoch values obtained from MUFFIT to a redshift-dependent plane of the form
\begin{equation}\label{eq:plane}
\mathrm{Age_M}+t_\mathrm{LB} (z)/\mathrm{Gyr} = a \cdot \log_{10} M_\star/\mathrm{M_\sun} + b \cdot \log_{10}r_\mathrm{c}/\mathrm{kpc} + c(z)\ ,
\end{equation}
where $a$ and $b$ are constants and $c(z)$ is a redshift-dependent function that we assume linear hereafter. As a sanity check, we fit Eq.~(\ref{eq:plane}) in each redshift bin in Figs.~\ref{fig:sizes_padova} to \ref{fig:sizes_bc03} and obtained that our formation epochs are properly fitted by a plane (typical differences below $0.5$~Gyr) and that $a$ and $b$ are compatible with no redshift evolution up to a $1$~$\sigma$ uncertainty level. From Eq.~(\ref{eq:plane}), the regions of the plane of constant formation epoch at a given redshift are those in which $M_\star \propto r_\mathrm{c}^\alpha$, where $\alpha=-b/a$. As the stellar population properties of this work are model dependent, we repeated this process for the sets of BC03 and EMILES SSP models. As a result, the regions of constant formation epoch are expressed by $\alpha=0.54 \pm 0.09$, $0.50 \pm 0.07$, and $0.55 \pm 0.10$ for BC03, EMILES+BaSTI, and EMILES+Padova00, respectively. The values of $\alpha$ are compatible within a $1$~$\sigma$ confidence level, even though the stellar population predictions from the different SSP models reveal quantitative discrepancies \citep[see e.g.][]{DiazGarcia2017a,DiazGarcia2017b}.

\begin{figure*}
\centering
\includegraphics[width=17cm]{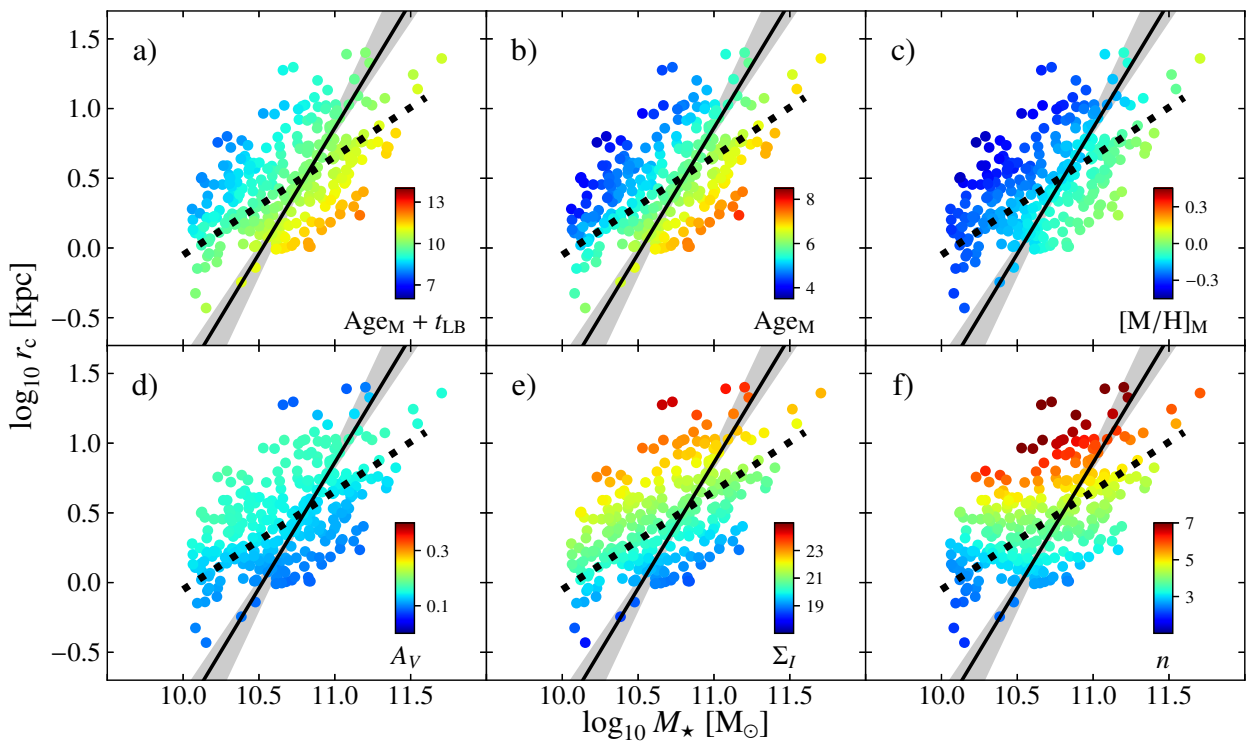}
\caption{Smoothed stellar population properties of quiescent galaxies for EMILES SSP models and Padova00 isochrones at $0.3 \le z < 0.5$ as a function of their positions in the stellar mass--size plane (coloured dots, see colour bar in each panel). The top panels show from left to right the mass-weighted formation epoch, age, and metallicity, and the bottom panels show extinction, luminosity surface brightness in the $I$ band, and S\'ersic index, respectively. To guide the eye, we illustrate the curve of constant formation epoch (black solid line) and its uncertainty (shaded area), as well as the stellar mass--size relation at this redshift bin (dotted black line).}
\label{fig:driver}
\end{figure*}

We checked whether other stellar population properties are also connected to the lines of constant formation epoch, that is, $M_\star \propto r_\mathrm{c}^\alpha$. We repeated the fitting process of Eq.~\ref{eq:plane}, but taking the values of mass-weighted age and metallicity instead of the formation epoch values. In Table~\ref{tab:alpha} we list the $\alpha$ coefficients for age and metallicity that we retrieved during this fitting. Figure~\ref{fig:driver} and Table~\ref{tab:alpha} illustrate that mass-weighted age and metallicity (panels b and c in Fig.~\ref{fig:driver}, respectively) agree with the $\alpha$ values obtained for the formation epochs. Owing to the large uncertainties for extinction, we constrained the line slope of constant extinction  to a range of $\alpha=0$--$2$. We reached the same conclusions independently of the SSP model set we used for the stellar population analysis.

\begin{table}
\caption{Slope $\alpha=-b/a$ of constant stellar population properties of quiescent galaxies in the stellar mass--size plane, see Eq.~(\ref{eq:plane}).}
\label{tab:alpha}
\centering
\begin{tabular}{lccc}
\hline\hline
&&& \\
 & EMILES & EMILES & \multirow{2}{*}{BC03}  \\
 & (Padova00) & (BaSTI) &  \\
\hline
&&& \\
Age$_\mathrm{M}+t_\mathrm{LB}$   & $0.55 \pm 0.10$  & $0.50 \pm 0.07$  & $0.54 \pm 0.09$  \\
$\mathrm{Age_M}$                 & $0.55 \pm 0.07$  & $0.50 \pm 0.09$  & $0.55 \pm 0.07$  \\
$\mathrm{[M/H]_M}$               & $0.43 \pm 0.13$  & $0.33 \pm 0.26$  & $0.34 \pm 0.25$  \\
&&& \\
\hline
\end{tabular}
\tablefoot{Stellar population parameters were obtained for the SSP model sets of \citet[][BC03]{Bruzual2003} and \citet[][EMILES]{Vazdekis2016}, the latter including BaSTI and Padova00 isochrones. From top to bottom, mass-weighted formation epoch, age, and metallicity.}
\end{table}

\subsection{Stellar mass--size relation of quiescent galaxies}\label{sec:driver_size}

It is well known that red galaxies exhibit a correlation between stellar mass and size, which also evolves across cosmic time \citep[e.g.][]{Shen2003,Trujillo2007,Buitrago2008,vanDokkum2008,vanderWel2014}. As an additional sanity check of the reliability of our sizes and stellar masses, we determined the correlation stellar mass--size and the size evolution of quiescent galaxies since $z\sim1$. This also allowed us to verify whether the lines of constant stellar population properties (age, formation epoch and metallicity) match the stellar mass--size relation of quiescent galaxies.

To parametrise the stellar mass--size relation, we used a power-law function of the form
\begin{equation}\label{eq:sms}
r_\mathrm{c}/\mathrm{kpc} = A \ \left( \frac{M_\star}{5\cdot 10^{10}\ \mathrm{M_\sun}} \right)^{\beta}\ , 
\end{equation}
where $A$ is the intercept of the stellar mass--size relation and $\beta$ its slope. Throughout this work, we assumed that $\beta$ is redshift independent, which has often been studied previously \citep[e.g.][]{Damjanov2011,Mclure2013,vanderWel2014}. However, the intercept $A$ evolves with redshift owing to the increase in size of quiescent galaxies. For simplicity, we assumed that the intercept is a linear function of redshift in the log-space, that is, $\log_{10}A(z) = \gamma \cdot z + \delta$. Making use of all the quiescent galaxies in our sample with stellar mass $M_\star \ge 10^{10.5}$~M$_\sun$ to avoid the flattening of the stellar mass--size relation at lower masses, we constrained the set of parameters $\beta$, $\gamma$, and $\delta$ that better fits Eq.~(\ref{eq:sms}). As there are discrepancies amongst the sample of quiescent galaxies depending on the SSP model \citep[see also][]{DiazGarcia2017a}, we retrieved the $\beta$, $\gamma$, and $\delta$ parameters for each SSP model set.

In Table~\ref{tab:sms} we summarise the parameters $\beta$, $\gamma$, and $\delta$ that best fit Eq.~(\ref{eq:sms}) for our sample. Although the sample of quiescent galaxies is not exactly the same for the three different sets of quiescent galaxies (see Table~\ref{tab:acs}), the parameters $\beta$ and $\gamma$ do not present a strong dependence on the SSP model set and they are compatible given the uncertainties. However, $\delta$ is more tightly related to the SSP model set, as expected owing to the quantitative differences between stellar content predictions from BC03 and EMILES \citep[see e.g.][]{DiazGarcia2017a}. The stellar mass--size correlation, $\beta$, is set around $0.72$ with an average error of $0.02$. The parameter $\gamma$ reflects the evolution in size of quiescent galaxies since $z\sim1$, for which we obtained a value of $\gamma = -0.43 \pm 0.02$. A negative value of $\gamma$ means that quiescent galaxies grow in size at lower redshifts, in  agreement with previous studies. Extrapolating these values, we obtained that the growth in size of quiescent galaxies from $z\sim1$ to $z=0$ is a factor of $2.5\pm0.1$. Regarding $\delta$, it presents a stronger dependence on the SSP model set used and values in the range $\delta = 0.6$--$0.8$ for EMILES and BC03 (see Table~\ref{tab:sms} for further details). In Fig.~\ref{fig:driver} we illustrate the stellar mass--size relation obtained for EMILES and Padova00 isochrones at $z \sim 0.4$ with a dotted line. 

Our results show that the lines of constant stellar population properties differ from the stellar mass--size relation (see dashed line in panels a to c in Fig.~\ref{fig:driver}). For extinction (panel d in Fig.~\ref{fig:driver}), we cannot confirm it given the uncertainties in our parameters. If they were equivalent, we would obtain that $\beta \sim 1/\alpha$. Nevertheless, the stellar content of quiescent galaxies is more tightly related to their position with respect to the stellar mass--size than to the stellar mass alone. This may be an indication that the assembly of the structures of galaxies (in this case represented by the effective radius) are related to their stellar content, but there are other mechanisms that also contribute to building it \citep[e.g.~mergers,][]{Naab2009}. It is worth mentioning that numerous studies mainly focus on the stellar mass--size relation separated red and blue galaxies based on morphology. Nevertheless, our sample is based on the intrinsic colours of galaxies, that is, on their stellar content. This implies that intrinsically red galaxies with late-type morphologies (S\'ersic indices lower than $2.5$) can be included in our sample (see Sect.~\ref{sec:driver_sersic}), which may explain discrepancies with other studies.

On the other hand, \citet[][]{vanderWel2014} studied the stellar mass--size relation of quiescent galaxies from the Cosmic Assembly Near-infrared Deep Extragalactic Legacy survey \citep[CANDELS, ][]{Grogin2011,Koekemoer2011}. These galaxies were selected using a $UVJ$ colour-colour diagram and the stellar masses were obtained using BC03 composite models. The authors found a slope of $\beta = 0.71$--$0.76$ since $z=1.25$, which is compatible with our results. In addition, after fitting the intercept obtained by \citet[][]{vanderWel2014} at $0.25 < z < 1.25$ to a lineal function with redshift, as in our work, we obtained that $\gamma=-0.46$ and $\delta=0.75$, which shows good agreement with our results when the BC03 SSP models were used. 

\begin{table}
\caption{Parameters describing the stellar mass--size relation of quiescent galaxies and its evolution since $z\sim1$, see also Eq.~(\ref{eq:sms}), and their uncertainties for BC03 and EMILES (including BaSTI and Padova00 isochrones) SSP models.}
\label{tab:sms}
\centering
\begin{tabular}{lccc}
\hline\hline
& \multirow{2}{*}{$\beta$} & \multirow{2}{*}{$\gamma$} & \multirow{2}{*}{$\delta$} \\
&&& \\
\hline
&&& \\
Padova00 & $0.71 \pm 0.02$ & $-0.41 \pm 0.02$ & $0.60 \pm 0.01$ \\
BaSTI    & $0.75 \pm 0.02$ & $-0.42 \pm 0.02$ & $0.57 \pm 0.01$ \\
BC03     & $0.68 \pm 0.02$ & $-0.46 \pm 0.02$ & $0.75 \pm 0.01$ \\
&&& \\
\hline
\end{tabular}
\end{table}


\subsection{Luminosity and mass surface densities in the stellar mass--size plane }\label{sec:driver_density}

Previous studies showed that the stellar content of red galaxies is more correlated with their luminosity or mass surface density ($\Sigma_{I}$ and $\Sigma_\mathrm{M}$, respectively) than with stellar mass \citep[e.g.][]{Kauffmann2003,Franx2008,Gargiulo2017}. To explore this topic, we studied the distribution of the surface brightness computed by \citet[][]{Griffith2012} for the ACS camera (defined as $\Sigma_{I} = I + 2.5 \log_{10} 2 \pi r_\mathrm{c}^2$) across the stellar mass--size plane. As for the stellar population properties derived from the ALHAMBRA photometry, we fit a plane to the original values of surface brightness for the $I$ band, which is the same band as we used to compute the effective radius (see panel e in Fig.~\ref{fig:driver}). We found that constant values of $\Sigma_I$ are properly expressed as $M_\star \propto r_\mathrm{c}^{2.4}$. This means that the regions of constant $\Sigma_I$ differ from the linear relation obtained in this work for constant formation epoch, age, and metallicity. The lines of constant $\Sigma_I$ better correlate with the stellar mass--size relation (see panel e in Fig.~\ref{fig:driver}). Interestingly, the stellar content of quiescent galaxies is more strongly correlated with $\Sigma_I$ than with the stellar mass.

The mass surface density is defined as $\Sigma_\mathrm{M} = M_\star / (2\pi r_\mathrm{c}^2)$, therefore constant values of $\Sigma_\mathrm{M}$ are along $M_\star \propto r_\mathrm{c}^2$. This trend is similar to the trend obtained for $\Sigma_{I}$, but this differs from the trend obtained for constant stellar population properties in Sect.~\ref{sec:driver_sp}, which shows a slope of $\alpha=0.50$--$0.55$.


\subsection{S\'ersic indices of quiescent galaxies in the stellar mass--size plane}\label{sec:driver_sersic}

Using S\'ersic indices as a proxy of morphology, we explored the correlation between morphology and stellar mass and size. We find that more than $80$~\% of the quiescent galaxies in our sample show a S\'ersic index, $n$, higher than $2.5$, which is the limit that is typically used in the literature to distinguish between early- and late-type galaxies. Quiescent galaxies with $n < 2.5$ typically populate the lower parts of the stellar mass--size plane, mainly in the low-mass regimes of our sample. This also illustrates a connection between quiescent and early-type morphologies.

For the case of the surface density, Sect.~\ref{sec:driver_density}, constant lines of the S\'ersic index for quiescent galaxies do not reflect the same correlations as are exhibited by constant stellar population parameters (see panel f in Fig.~\ref{fig:driver}). Nevertheless, this correlation follows similar trends as the surface density and the stellar mass--size relation.


\subsection{Velocity dispersions and the driver of the stellar populations of quiescent galaxies}\label{sec:driver_velocity}

In the past decades, the tight correlation between stellar populations and velocity dispersion (usually quoted at one effective radius, $\sigma_\mathrm{e}$) obtained in several spectroscopic studies has increasingly be considered evidence that the velocity dispersion drives the evolution of galaxies \citep[$\sigma$; e.g.][]{Trager2000,Gallazzi2006,Graves2010,Cappellari2013}. Because our sample of quiescent galaxies was obtained through photometric data (resolving power $R\sim20$ for ALHAMBRA), we cannot provide an estimate of $\sigma_\mathrm{e}$ for each of our galaxies. Instead, we built distributions of $\sigma_\mathrm{e}$ in the stellar mass--size plane from spectroscopic quiescent galaxies at the same redshift range as our sample. 

\begin{figure}
\resizebox{\hsize}{!}{\includegraphics{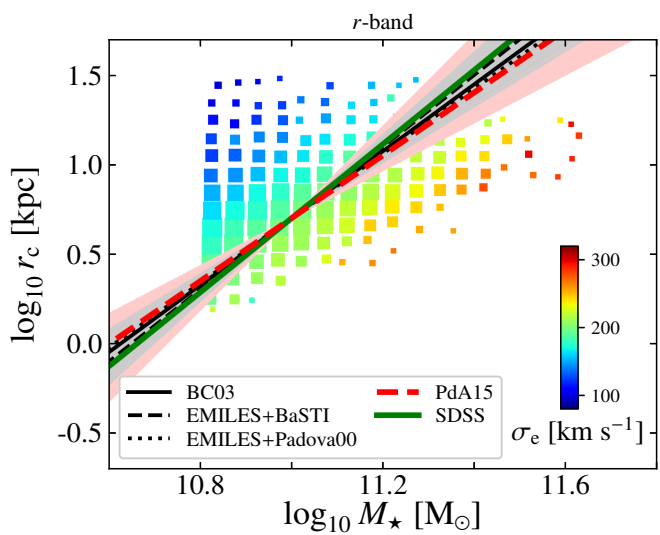}}
\caption{Average values of the velocity dispersion in the stellar mass--size plane for quiescent galaxies from the SDSS at $0.02 \le z \le 0.08$ for the $r$ band. The size of the marker illustrates the number of galaxies in each bin, and its colour illustrates the average velocity dispersion at one effective radius (see colour bar). Black lines show the empirical relation of constant formation epochs (Eq.~(\ref{eq:plane})) obtained for BC03, EMILES+BaSTI, and EMILES+Padova00 SSP models (solid, dashed, and dotted black lines, respectively). Curves of constant velocity dispersions for one effective radius using SDSS data and the results from \citet{PeraltaArriba2015} are illustrated with green and red lines, respectively. For each case, the shaded areas show the uncertainties in the fits of constant galaxy properties.}
\label{fig:sigma}
\end{figure}

Making use of the NYU Value Added Galaxy Catalogue DR7 (NYU catalogue hereafter) of the Sloan Digital Sky Survey \citep[SDSS, ][]{Blanton2005,Blanton2007,AdelmanMcCarthy2008,Padmanabhan2008}, we built a sample of quiescent galaxies with spectroscopic measurements of $\sigma$, stellar mass, and photometric effective radius for the $r$ and $i$ bands. Adopting the rest-frame colour limits proposed by \citet[][]{Schawinski2014}, we selected quiescent galaxies at $0.02 \le z \le 0.08$ from the NYU catalogue using the k-corrected $u$, $g$, and $J$ magnitudes \citep[$ugJ$ colour--colour diagram, see Fig.~4 in][]{Schawinski2014}. The stellar mass was scaled to a \citet[][]{Chabrier2003} IMF, and the sample of quiescent galaxies was constrained to $\log_{10}M_\star \ge 10.8$ for the following reasons: (i) to be complete in stellar mass; (ii) to avoid the flattening of the stellar mass--size relation at $\log_{10}M_\star \sim 10.6$; and (iii) to match the stellar mass range in which our sample is complete since $z\sim1$. As in \citet[][]{PeraltaArriba2014}, we applied additional restrictions to the NYU catalogue to avoid unreliable data: $70 \le \sigma \le 320$~km~s$^{-1}$, $0.3 \le r_\mathrm{e} \le 30$~kpc, and $r_\mathrm{e}\ge 1$~\arcsec. The NYU catalogue contains sizes for $r$ and $i$ bands, we therefore built two final samples of $\sim11\,600$ quiescent galaxies after applying the above restrictions. Owing to the aperture effects of SDSS, the velocity dispersions of quiescent galaxies from the NYU catalogue are out to one $r_\mathrm{e}$. We therefore corrected them for aperture effects using Eq.~1 of \citet[][]{Cappellari2006}. In Fig.~\ref{fig:sigma} we illustrate the mean $\sigma_\mathrm{e}$ of the final SDSS sample at different regions across the stellar mass--size plane. At fixed stellar mass, more compact quiescent galaxies from the SDSS exhibit larger $\sigma_\mathrm{e}$ than their more extended counterparts. As in Sect.~\ref{sec:driver_sp}, we fit the distribution of individual spectroscopic galaxies to a plane (see Eq.~(\ref{eq:plane})) and obtained an analytic expression of constant $\sigma_\mathrm{e}$. As a result of the fitting, constant values of $\sigma_\mathrm{e}$ are properly described by $M_\star \propto r_\mathrm{c}^{0.48\pm0.01}$ (see the green line in Fig.~\ref{fig:sigma}) with typical differences with respect to the plane below $26$~km~s$^{-1}$. The parameter $\sigma_\mathrm{e}$ is not linked to a SED-fitting analysis based on colours, meaning that it is independent of our predictions of stellar population properties. We did not estimate dynamical masses (usually based on the virial theorem and homology), but stellar masses derived from colours and luminosities. This result reflects that at least in the nearby Universe, constant values of formation epoch, age, and metallicity across the stellar mass--size plane occupy regions of constant velocity dispersion within the uncertainties (see the black lines in Fig.~\ref{fig:sigma}). This result is consistent independently of the band that is used to determine projected sizes, that is, for $r$ and $i$ bands.

At intermediate redshifts, $z\sim 1$, fewer spectroscopic studies across the stellar mass--size plane that include reliable $\sigma_\mathrm{e}$ measurements are available. However, \citet{PeraltaArriba2014,PeraltaArriba2015} built samples of massive spheroid-like galaxies \citep[$\log_{10}M_\star \gtrsim 10.8$ for an IMF of ][$n\ge 4$]{Chabrier2003} at $z\sim1$. In particular, \citet[][]{PeraltaArriba2015} built a sample to cover a broad range on the stellar mass--size plane, which fits the range in stellar mass and redshift that we explore with the quiescent galaxies of ALHAMBRA. \citet[][]{PeraltaArriba2015} reported a breakdown in the homology assumption \citep[also confirmed in][]{PeraltaArriba2014}, yielding that constant $\sigma_\mathrm{e}$ values satisfies $M_\star \propto r_\mathrm{c}^{0.57\pm0.18}$ (see the dashed red line in Fig.~\ref{fig:sigma}), which is compatible with no redshift evolution since $z\sim1$. This result is in agreement with those obtained from the SDSS for the nearby Universe, but also with the constant regions of stellar population properties of quiescent galaxies since $z\sim1$ (see Fig.~\ref{fig:sigma}).

\subsection{Main driver of the stellar content of galaxies}\label{sec:driver_how}

Some indications support that velocity dispersions within one effective radius are a good candidate to be tightly correlated with the stellar content of quiescent galaxies, even more than the stellar mass or the other parameters we explored. Nevertheless, this correlation between observables does not confirm that the velocity dispersion drives the evolution of stellar populations. There might be an underlying mechanism favouring the creation of stars that also increases the velocity dispersion. However, this would reflect that the driver of the stellar populations of galaxies would be partly linked to the dynamical properties of galaxies, as well as to their gravitational potential.

In a similar way, \citet[][]{Barone2018} explored plausible drivers of the stellar content of $625$ early-type galaxies using Integral Field Spectroscopy (IFS) from the Sydney-Australian-Astronomical-Observatory Multi-object Integral-Field Spectrograph galaxy survey \citep[SAMI, ][]{Bryant2015}. The authors compared the correlations between dynamic masses ($M_\mathrm{D}$), gravitational potential ($\Phi$), surface density ($\Sigma$) and ages, metallicities, and $\alpha$ abundances. Under the assumption of homology in virial equilibrium, we would obtain that $M_\mathrm{D}=K\sigma_\mathrm{e}^2 r_\mathrm{e}/G$, $\Phi \propto M_\mathrm{D}/r_\mathrm{e} \propto \sigma^2$ and $\Sigma \propto M_\mathrm{D}/r_\mathrm{e}^2 \propto \sigma^2/r_\mathrm{e}$. Based on the correlation between the stellar population parameters and $M_\mathrm{D}$, $\Phi,$ and $\Sigma$, \citet[][]{Barone2018} reported that the metallicity presents a stronger correlation with $\Phi$ and the age with $\Sigma$. Using the stellar population parameters of quiescent galaxies from ALHAMBRA (Sect.~\ref{sec:sp_size} and Figs.~\ref{fig:sizes_padova} to \ref{fig:sizes_bc03}), we explored whether stellar populations are more strongly correlated with gravitational potential ($\Phi_\mathrm{M}$) or surface density ($\Sigma_\mathrm{M}$). In our case, we used their non-dynamical definitions, that is, $\Phi_\mathrm{M} \propto M_\star/r_\mathrm{c}$ and $\Sigma_\mathrm{M} \propto M_\star/r_\mathrm{c}^2$. Similarly to \citet[][]{Barone2018}, we quantified the correlation between stellar population parameters and $\Phi_\mathrm{M}$ and $\Sigma_\mathrm{M}$ through a Spearman correlation coefficient, and we quantified their scatter assuming a linear relation. As a result, all the stellar population parameters explored in this research exhibit larger correlation coefficients for $\Phi_\mathrm{M}$ than for $\Sigma_\mathrm{M}$ at any stellar mass and redshift. It is also remarkable that the scatter with $\Phi_\mathrm{M}$ is lower than for $\Sigma_\mathrm{M}$. This is expected because we observationally constrained that constant values of formation epoch, age, and metallicity lie along linear relations of the form $M_\star \propto r_\mathrm{c}^{0.5}$, which is more similar to the gravitational potential (i.e.~$M_\star \propto r_\mathrm{c}$) than to the surface density (i.e.~$M_\star \propto r_\mathrm{c}^2$). The observational discrepancies with respect $\Phi$ and $\Sigma$ can be an indication that other key physical factors may also drive the stellar assembly (e.g.~internal kinematics or the primordial angular momentum of gas clouds), external mechanisms such as mergers \citep[e.g.][]{Khochfar2006b,Naab2006,Perea2016,Solanes2016}, or the masses of dark matter haloes \citep[e.g.][]{Huang2018}, which should be studied in detail in future works.

Interestingly, \citet[][]{Mcdermid2015} and \citet[][]{Li2018} retrieved similar results from spectroscopic studies using the dynamical mass instead of the stellar mass. The authors also found that curves of constant velocity dispersion across the dynamical mass--size plane showed the same trend as constant age, metallicity, and [$\alpha$/Fe]. This would support other recent studies claiming that dynamical and stellar mass are proportional to  a factor that does not significantly change at different stellar mass range \citep{Gavazzi2007,Taylor2010,Zahid2017}.


\section{Growth in size of quiescent galaxies}\label{sec:discussion_grow}

The analysis conducted in this work reveals that more compact quiescent galaxies are older, more metal rich, and less reddened by dust than their more extended counterparts at fixed stellar mass. These results allow us to shed light on the different mechanism acting in the growth in size of these galaxies since $z\sim 1$. The correlation between size and age for quiescent galaxies is additional evidence to discard the puffing-up scenario as causing the growth in size of galaxies because this mechanism would imply that compact galaxies were younger than their more extended counterparts.

Mergers and the progenitor bias agree with the results obtained in this topic. Both mechanisms have been extensively studied to explain the growth in size of quiescent galaxies \citep[][]{Hopkins2009,Naab2009,Valentinuzzi2010,Trujillo2011,Carollo2013,Belli2015}. The latter because late-types or star-forming galaxies are typically more extended than early-type or quiescent galaxies, and they therefore lie on the upper parts of the quiescent stellar mass--size relation after their star formation has shut down. Recent studies revealed evidence for remnants of low star formation activity in quiescent galaxies \citep[e.g.][]{Vazdekis2016}, which is usually referred to as `frosting'. Nevertheless, it is not clear how this may affect the galaxy size. If the remnants of star formation lie in the inner parts of the galaxy, frosting would reduce the apparent size of the galaxy. Conversely, new stars created in the surroundings of the inner parts (such as a disc or shell around the bulge or galactic nucleus) would probably produce an apparent growth in size for the host galaxy. Therefore frosting should not be discarded as another feasible mechanism to produce the increase in size of galaxies. However, the importance of this mechanism is also related to the star formation activity, which is usually low and should be explored as a function of cosmic time. Overall, the growth in size of quiescent galaxies may not be driven by a unique mechanism.

As revealed by \citet[][]{DiazGarcia2013}, since $z\sim1$ the merger history of galaxies does not depend on the galaxy size and the growth in size through mergers seems to similarly affect the whole quiescent population. Under this scenario, mergers would contribute to eliminate compact quiescent galaxies throughout cosmic time. A detailed study of the evolution of the number density of compact quiescent galaxies would help to shed light on the importance (or lack thereof) of mergers in terms of the size evolution of quiescent galaxies. Unfortunately, there is no clear consensus about the evolution of the number density of compact quiescent galaxies because some previous studies reported that this number is roughly constant \citep{Saracco2010,Damjanov2014,Damjanov2015,Gargiulo2017}, whereas other work stated that it exhibits a significant or mild decrement \citep[][]{Valentinuzzi2010,Cassata2013,Poggianti2013,Quilis2013,Trujillo2014,vanderWel2014}. For instance, if the co-moving number density of compact quiescent galaxies were lower at more recent epochs, this would reflect that mergers occur, although their importance would be hard to determine. On the other hand, an increasing number of compact galaxies would reflect an arrival of new compact quiescent galaxies, that is, the progenitor bias \citep{vanDokkum2001} would also alter the stellar mass--size relation. The sample of quiescent galaxies with reliable size measurements we presented here is not large enough for a reliable and direct study of the evolution in number density of compact quiescent galaxies.

From the results obtained by \citet[][]{DiazGarcia2013,DiazGarcia2017b}, it is possible to draw the following general picture on the growth in size of the quiescent population. Firstly, instead of a unique mechanism to explain the growth in size of quiescent galaxies, there seems to be a combination of various mechanisms acting in parallel to produce the increase in size observed in quiescent galaxies: mergers, the progenitor bias and possibly frosting. In a downsizing scenario, the first galaxies that quenched their star formation were the galaxies that experienced the most efficient star formation episodes,  with very short formation timescales: These are in turn the most massive galaxies. The formation and evolution of the stellar populations of quiescent galaxies also relate to the size of the galaxy, that is, to the stellar mass density of the galaxy, where the most compact galaxies at fixed stellar mass would be formed earlier and more efficiently. After their star formation ceases, they were incorporated into the quiescent population, and under this hypothesis, all the galaxies that were formed later will present more extended sizes and younger stellar populations, which produces the size--age correlation. In addition, there is evidence that star-forming galaxies typically contain stellar populations with lower metallicities than the quiescent ones \citep[e.g.][]{Gallazzi2014,Peng2015}. Consequently, the star-forming galaxies quenching their star formation activity would populate the upper parts of the stellar mass--size plane and they would be able to exhibit lower metallicity than the already quenched galaxies. However, some authors argued that the metallicity changes during the quenching phase \citep[see e.g.][]{Peng2015}, and depending on the mechanism that shuts the star formation activity down, the progenitor bias may not contribute to the size--metallicity correlation. A further study is necessary to confirm this, however. When these galaxies belong to the quiescent sample, some of them would continue to grow in size through mergers and (maybe) frosting. This means that the continuous arrival of new quiescent galaxies with larger sizes may be an important mechanism that would contribute to modify the stellar mass--size relation \citep[e.g.~][ reported that the progenitor bias can only be responsible for about half of the growth in size at $1<z<1.6$]{Belli2015}.

Secondly, although the growth in size is likely produced by a combination of various mechanisms, their role would depend on cosmic time. For instance, massive quiescent galaxies ($\log_{10}M_\star \ge 11.2$) suffer a rapid increase in number at $1<z<3$ \citep[e.g.][]{Arnouts2007,DominguezSanchez2011,Ilbert2010,Ilbert2013}, which implies a more striking contribution of the progenitor bias than at $z<1$, provided that the majority of massive galaxies are already red at this redshift range \citep[e.g.][]{Davidzon2013,Ilbert2013,Moustakas2013,Tomczak2014}. In addition, major mergers were more frequent at higher redshifts \citep[e.g.][]{Lin2008,deRavel2009,LopezSanjuan2012,Xu2012,LopezSanjuan2013}, and therefore the contribution of mergers to the increase in size might be more relevant. These facts would contribute to explain that the increase in size of quiescent galaxies were more efficient at $1<z<2$ rather than at $0<z<1$ \citep[e.g.][]{Trujillo2007,vanDokkum2008,Newman2012,vanderWel2014}.

Thirdly, at fixed redshift, the role of the different mechanisms to produce the growth in size of quiescent galaxies may vary with stellar mass. As discussed above, at $z<1$ there is a generalised increase in the number density of quiescent galaxies that at decreasing stellar mass is systematically more prominent \citep[see e.g.][]{DiazGarcia2017b,Damjanov2019}, which means that the importance of the progenitor bias is different at different stellar mass ranges.


\section{Comparison with previous studies}\label{sec:sizes_previous}

Since the discovery of the stellar mass--size relation, several previous works have studied the stellar content of compact galaxies at different ranges of redshift and stellar mass \citep{Shankar2009,Trujillo2011,Belli2015,Mcdermid2015,Fagioli2016,Gargiulo2017,Scott2017,Williams2017,Li2018,Wu2018,Damjanov2019}. We here extended this type of studies up to $z\sim1$ to also explore correlations with metallicity and extinction. We devote this section to comparing our results, obtained using photometric data, with the spectroscopic work in the literature (see Table~\ref{tab:size_pw} for a brief summary).

\begin{table*}
\caption{Brief description of samples used in previous work to determine stellar population parameters of galaxies as a function of size.}
\label{tab:size_pw}
\centering
\begin{tabular}{lcccc}
\hline\hline
\multicolumn{1}{c}{\multirow{2}{*}{References}} & \multicolumn{1}{c}{\multirow{2}{*}{Redshift}} & \multicolumn{1}{c}{\multirow{2}{*}{Stellar mass}} & \multirow{2}{*}{Number} & \multirow{2}{*}{Parameters}\\
&&&& \\
\hline
&&&& \\
\citet{Shankar2009}$^\dag$  & $0.01 < z < 0.3$   & $10^{\ 9.7} < L_r^\mathrm{corr} < 10^{11.5}$  & $\sim48\,000$ & Age        \\
\citet{Trujillo2011} & $z<0.1$ and $z\sim1$   & $M_\star \gtrsim 10^{10}$             & $2656$ and $228$  & Age                                \\
\citet{Belli2015}    & $1.0\ \ < z < 1.6$ & $M_\star > 10^{10.6\ }$               & $62$        & Age                                \\
\citet{Mcdermid2015} & $z \lesssim 0.01$  & $10^{\ 9.8} < M_\star < 10^{12\ \ }$  & $260$       & Age, [M/H], [$\alpha/\mathrm{Fe}$] \\
\citet{Fagioli2016}  & $0.2 < z < 0.8$    & $10^{10.5} < M_\star < 10^{11.5}$     & $1\,519$    & Age                                \\
\citet{Gargiulo2017} & $0.5 < z < 1.0$    & $M_\star > 10^{11\ \ }$               & $>2\,000$   & Age                                \\
\citet{Scott2017}    & $0.02 < z < 0.06$  & $M_\star \gtrsim 10^8$                & $1\,319$    & Age, [M/H], [$\alpha/\mathrm{Fe}$] \\
\citet{Williams2017} & $z\sim 1.2$        & $M_\star > 10^{11\ \ }$               & $55$        & Age                                \\
\citet{Li2018}       & $z < 0.1$          & $M_\star \gtrsim 10^{9.5}$            & $952$       & Age, [M/H], $M_\star/L$            \\
\citet{Wu2018}       & $ 0.6 \le z \le 1.0$     & $M_\star \gtrsim 10^{10}$     & $467$    & Age
\\
\citet{Damjanov2019} & $0.1 < z < 0.6$          & $M_\star \gtrsim 10^{10}$    & $\sim3\,500$  & Age
\\
&&&& \\
\hline
\end{tabular}
\tablefoot{From left to right, reference of the work, redshift bin, stellar mass range, number of galaxies, and stellar population parameters  of each sample. All the stellar masses (luminosities) are in solar units [M$_\sun$] ([L$_\sun$]). All the work involved spectroscopic data. \citet{Belli2015} and \citet{Gargiulo2017} also included photometric SED-fitting to complement spectroscopic predictions. \\ ($\dag$) \citet{Shankar2009} used luminosities as stellar mass proxy.}
\end{table*}


\subsection{Stellar populations of compact and extended galaxies in the nearby Universe}

Early attempts to unveil the discrepancies between the stellar populations of compact and extended galaxies were carried out using spectroscopic data from the SDSS.  \citet{Shankar2009} explored the relation between half-light radii $r_\mathrm{e}$ and the ages of early-type galaxies from the SDSS at $0.013 < z < 0.3$. The authors revealed that early-type galaxies of ages older than $9$~Gyr systematically present lower effective radii of $\Delta \log_{10}r_\mathrm{e}\sim 0.1$~dex than those younger than $6$~Gyr. This agrees qualitatively with our results.

The pioneering work by \citet{Trujillo2011} determined systematic discrepancies between the extreme age quartiles of early-type galaxies at fixed dynamical mass, where the young galaxies exhibited larger effective radii and a correlation with velocity dispersion. \citet{Trujillo2011} extended this result at $z\sim1$ where little variation in the segregation of the age within the stellar mass--size relation was found, which implies that the evolution of this relation is independent of the galaxy age. This strongly constrained the puffing-up mechanism in explaining the growth in size of galaxies and agrees with our results: the puffing-up model is inconsistent with the observed distribution of ages across the stellar mass--size. However, \citet{Trujillo2011} found smaller age differences between the extreme quartiles than we did. This may be a consequence of the differences in the sample selection because \citet{Trujillo2011} analysed early-type galaxies instead of colour-based quiescent galaxies.

By means of integral field spectroscopy, \citet{Mcdermid2015} obtained similar correlations across the stellar mass--size plane for index-based ages and metallicities and the mass-weighted ones: more compact galaxies present both older ages and higher metallicities. This result agrees with our predictions of quiescent galaxies at $0.1 < z < 0.3$ and $\log_{10}M_\star > 10$. Whilst our sample was built using the intrinsic colours of galaxies as spectral discriminator, the sample of \citet{Mcdermid2015} was morphologically selected and preserves the same correlations with size that we obtained, in part because early-type galaxies frequently present red colours. From the combined flux of spatially resolved galaxies within one effective radius, \citet[][]{Scott2017} and \citet[][]{Li2018} also confirmed the segregation of the stellar population properties with size, that is, more compact galaxies are older and more metal rich. As we also obtain in Sect.~\ref{sec:driver_velocity}, \citet[][]{Li2018} showed that the distribution of velocity dispersion changes according to the stellar population parameters of galaxies for stellar masses of $\log_{10}M_\star \gtrsim 10.4$.  \citet[][]{Li2018} revealed a correlation between the size and the gradients of metallicity and age, which we cannot compare with our results because our study involves integrated stellar populations. Interestingly, \citet{Mcdermid2015} and \citet{Scott2017} reported more alpha-enhanced stellar populations for compact early-type galaxies, although we are not able to compare this correlation.


\subsection{Correlation between age and size at moderate redshift}

At moderate redshift, previous spectroscopic studies have mainly focused on the ages as a function of size, but metallicity remains not strongly determined. \citet{Belli2015} explored the stellar populations of quiescent galaxies at $1 < z < 1.6$ and obtained that older quiescent galaxies populate the lower parts of the stellar mass--size diagram, which is in agreement with our results at intermediate redshift. By use of the star formation history, \citet{Belli2015} estimated that extended quiescent galaxies arrived later in the quiescent sample than the compact galaxies, which were mainly formed in later formation epochs than the compact counterpart. In addition, \citet{Belli2015} constrained the importance of the progenitor bias in the assembling of the stellar mass--size of quiescent galaxies, which would partly explain the age--size correlation that we observe at lower redshifts.

At a similar redshift range as we explored here, \citet{Fagioli2016} and \citet[][]{Gargiulo2017} also found a general size--age correlation at intermediate redshift as well. However, \citet{Fagioli2016} did not find evidence of a correlation between age and size for quiescent galaxies more massive than  $\log_{10}M_\star = 11$.  \citet[][]{Gargiulo2017} complemented their study by including the co-moving number densities of compact and extended quiescent galaxies. As a result, extended quiescent galaxies are younger than their compact counterparts since $z=0.8$ according to a non-passive evolution \citep[see also][]{DiazGarcia2017b}.

At $0.1 < z < 0.6$, \citet[][]{Damjanov2019} analysed the distributions of the age-sensitive estimator $D_\mathrm{n}4000$ by segregating a sample of quiescent galaxies according to stellar mass and redshift. For quiescent galaxies at $\log_{10}M_\star < 10^{11}$, the $D_\mathrm{n}4000$ distributions of younger quiescent galaxies (defined as $1.5 < D_\mathrm{n}4000 < 1.75$) showed larger sizes than their older counterpart ($D_\mathrm{n}4000 > 1.75$), meaning a correlation between age and size in which compact galaxies are younger. However, as in \citet{Fagioli2016}, galaxies more massive than $\log_{10}M_\star=10^{11}$ do not show a trend in the distributions of $D_\mathrm{n}4000$ with size. In a similar way and since $z\sim1.2$, \citet{Williams2017} also measured the $D_\mathrm{n}4000$ index for different size bins, but the authors also complemented their results by including age predictions from the {H$\delta$} index. For {H$\delta$}, \citet{Williams2017} estimated age differences of $\sim2.5$~Gyr between compact and extended sources, but for $D_\mathrm{n}4000$ this difference decreases to $\sim0.3$~Gyr. This supports our difference in age of $\sim1$~Gyr between compact and extended quiescent galaxies. Also interpreting age-sensitive index variations, $D_\mathrm{n}4000$ and H$\delta$, \citet[][]{Wu2018} discerned younger stellar populations in quiescent galaxies at larger sizes and fixed stellar mass. More precisely, larger quiescent galaxies are $\sim500$~Myr younger than the compact ones, which is a smaller differences in age than our results ($\sim1$~Gyr). Nevertheless, we used mass-weighted stellar population properties. This may introduce discrepancies with respect to the outcomes obtained using luminosity-weighted properties because composite stellar population models can be more affected by recent bursts of star formation \citep[][]{Ferreras2004,Serra2007,Ferreras2009,Rogers2010}. It is of note that \citet[][]{Wu2018} obtained consistent results using three independent methods to define their quiescent galaxy sample.

The results from different studies were obtained by different techniques based on other sensitive features, SSP models, star formation histories, and even with different criteria to separate quiescent galaxies, which may yield quantitative discrepancies and/or biased samples. Despite these potential discrepancies, there is an overall agreement and our work converges to a common qualitative outcome with the results obtained from other spectroscopic studies.


\section{Summary and conclusions}\label{sec:sizes_conclusions}

After selecting all the quiescent galaxies from ALHAMBRA with reliable sizes (circularised effective radius) in HST fields, we obtained a sub-sample of more than $800$ of galaxies with stellar population parameters. We used this sub-sample to discern any correlations between the stellar content of quiescent galaxies and their sizes at fixed stellar mass. 

Our results show evidence for strong correlations between the size and the age, formation epoch, and metallicity of quiescent galaxies, although there are indications for a slight correlation with extinction as well. This outcome does not depend on the SSP model used, and it was obtained for three SSP model predictions: BC03, EMILES with BaSTI isochrones, and EMILES with Padova00 isochrones. 

These correlations were studied in detail and revealed remarkable insights. Firstly, the size of a quiescent galaxy and its age are tightly correlated. A generalised trend is that more compact quiescent galaxies are older than their more extended counterparts. The difference in age is established around $1$~Gyr for stellar masses $\log_{10}M_\star > 9.6$ and up to $z\sim0.9$, where the average size--age correlation is $\langle \bar{m}_\mathrm{Age_M} \rangle \sim -1.4 \pm 0.2$ (see Eq.~(\ref{eq:size_sp_par})). This trend is reflected in the dust-corrected luminosity at $2\,800$~\AA, where less compact quiescent galaxies exhibit higher ultraviolet luminosities, possibly as a result of a more recent or longer star formation activity. Secondly, compact quiescent galaxies are systematically more metal rich than the less dense galaxies at the same stellar mass. Differences between metallicity of compact and extended quiescent galaxies amount to $\sim0.07$~dex, at least since $z=0.9$, where the correlation size--metallicity is $\langle \bar{m}_\mathrm{[M/H]_M} \rangle \sim -0.11 \pm 0.02$ on average. Finally, extended quiescent galaxies may present slight higher extinctions of $<0.1$ independently of their stellar mass range and redshift, $\langle \bar{m}_{A_V} \rangle \sim 0.03 \pm 0.01$.

We determined the regions  of constant formation epoch  across the stellar mass--size plane to empirically set constraints on the physical mechanism  that govern the evolution of stellar populations in galaxies. Our results indicate that this relation is properly expressed by $M_\star \propto r_\mathrm{c}^\alpha$, for which we obtained values of $\alpha=0.54 \pm 0.09,\ 0.50 \pm 0.07,\text{and}\ 0.55 \pm 0.10$ for BC03 and EMILES SSP models, respectively (the latter including BaSTI and Padova00 isochrones). This is also supported by other stellar population parameters such as mass-weighted age and metallicity. After studying several stellar population properties, some indications support the idea that constant velocity dispersions within one effective radius lie in constant regions of stellar population properties across the stellar mass--size plane, that is, they are tightly correlated with the stellar content of quiescent galaxies. This may reflect that the driver of the stellar populations of galaxies might be partly linked to the dynamical properties of galaxies, as well as to their gravitational potential.

In view of these results, the puffing-up scenario can be discarded as a responsible mechanism of the growth in size of galaxies \citep[in good agreement with the pioneering work by][which gave strong evidence against the puffing-up mechanism of gas expulsion]{Trujillo2011} because this scenario would imply that compact galaxies were younger than their more extended counterparts. However, the progenitor bias agrees with the results obtained in this research, as well as mergers. The progenitor bias and mergers homogeneously act across the stellar mass--size plane, which may contribute to explain the growth in size of quiescent galaxies since $z\sim1$, where the importance of these mechanisms may vary depending on the redshift and stellar mass range explored.

Finally, we compared our results with previous spectroscopic work on this topic, mostly focused on probing potential age differences between compact and extended galaxies. Overall, there is a very good agreement regarding these age effects. Interestingly, our results greatly extend some of the conclusions obtained in the nearby Universe: more compact quiescent galaxies are more metal rich than their extended counterparts, but this is also observed for the first time since $z\sim 1$ through data from the large-scale multi-filter ALHAMBRA survey. A correlation between size and extinction was also explored, which constitutes a complement for future studies.


\begin{acknowledgements}

This work has been supported by the Programa Nacional de Astronom\'ia y Astrof\'isica of the Spanish Ministry of Economy and Competitiveness (MINECO, grants AYA2012-30789 and AYA2015-66211-C2-1-P), by the Government of Arag\'on (Research Group E103), by the Ministry of Science and Technology of Taiwan (grant MOST 106-2628-M-001-003-MY3), and by the Academia Sinica (grant AS-IA-107-M01). The authors also acknowledge support from the Spanish Ministry for Economy and Competitiveness and FEDER funds through grants AYA2010-15081, AYA2010-15169, AYA2010-22111-C03-01, AYA2010-22111-C03-02, AYA2011-29517-C03-01, AYA2012-39620, AYA2013-40611-P, AYA2013-42227-P, AYA2013-43188-P, AYA2013-48623-C2-1, AYA2013-48623-C2-2, ESP2013-48274, AYA2014-57490-P, AYA2014-58861-C3-1, AYA2016-76682-C3-1-P, AYA2016-76682-C3-3-P, AYA2016-77846-P Generalitat Valenciana projects Prometeo 2009/064 and PROMETEOII/2014/060, Junta de Andaluc\'{\i}a grants TIC114, JA2828, P10-FQM-6444, and Generalitat de Catalunya project SGR-1398. 

In this work we made use of the NYU Value Added Galaxy Catalogue DR7, which includes data from the Sloan Digital Sky Survey (SDSS) and the Two Micron All Sky Survey (2MASS). Funding for the SDSS has been provided by the Alfred P.~Sloan Foundation, the Participating Institutions, the National Aeronautics and Space Administration, the National Science Foundation, the U.S.~Department of Energy, the Japanese Monbukagakusho, and the Max Planck Society. The SDSS Web site is \url{http://www.sdss.org/}. The SDSS is managed by the Astrophysical Research Consortium (ARC) for the Participating Institutions. The Participating Institutions are The University of Chicago, Fermilab, the Institute for Advanced Study, the Japan Participation Group, The Johns Hopkins University, Los Alamos National Laboratory, the Max-Planck-Institute for Astronomy (MPIA), the Max-Planck-Institute for Astrophysics (MPA), New Mexico State University, University of Pittsburgh, Princeton University, the United States Naval Observatory, and the University of Washington. This publication makes use of data products from the 2MASS, which is a joint project of the University of Massachusetts and the Infrared Processing and Analysis Center/California Institute of Technology, funded by the National Aeronautics and Space Administration and the National Science Foundation. 

The authors thank the constructive and fruitful comments provided by the anonymous referee that surely helped to improve this paper.

Throughout this research, we made use of the \texttt{Matplotlib} package \citep{Hunter2007}, a 2D graphics package used for \texttt{Python} that is designed for interactive scripting and quality image generation. This paper is dedicated to Marian Le\'on Canalejo for being there when L.A.D.G.~needed her most and for her patience and continuous encouragement while finishing his Ph.D.

\end{acknowledgements}


\bibliographystyle{aa}
\bibliography{ms}


\end{document}